\documentclass[aps,nofootinbib,notitlepage,11pt]{revtex4-1}
\usepackage{slashed}
\usepackage{amsmath,amsfonts,amssymb}
\usepackage{graphicx}
\usepackage{bm}
\usepackage[colorlinks,linkcolor=blue,citecolor=blue]{hyperref}
\usepackage[usenames,dvipsnames,svgnames,table]{xcolor}

\newcommand{\di}{\text{d}}
\newcommand{\be}{\begin{equation}}
\newcommand{\ee}{\end{equation}}
\newcommand{\bea}{\begin{eqnarray}}
\newcommand{\eea}{\end{eqnarray}}
\newcommand{\ba}{\begin{eqnarray}}
\newcommand{\ea}{\end{eqnarray}}

\newcommand{\Tr}{\mbox{Tr}\;}

\begin{document}
\title{Polarization of the $\phi$ meson  in the hadronic phase with nucleon scatterings and a viscous hydrodynamic background}

\author{Eduardo Grossi}
\email{eduardo.grossi@unifi.it }
\affiliation{Dipartimento di Fisica, Universit\`a di Firenze and INFN Sezione di Firenze, via G. Sansone 1,
50019 Sesto Fiorentino, Italy}

\author{Andrea Palermo}
\email{andrea.palermo@stonybrook.edu}
\affiliation{Center for Nuclear Theory, Department of Physics and Astronomy, Stony Brook University, Stony Brook, New York 11794--3800, USA}

\author{Ismail Zahed}
\email{ismail.zahed@stonybrook.edu}
\affiliation{Center for Nuclear Theory, Department of Physics and Astronomy, Stony Brook University, Stony Brook, New York 11794--3800, USA}

\begin{abstract}
We extend our previous work on the spin alignment of the $\phi$ vector meson in the hadronic phase of the Quark Gluon Plasma, by including effects of nucleon scatterings. The emission rates are calculated in a realistic hydrodynamic background simulated with the code Fluid$u$m, for different beam energies. We find that all the effects taken into account cannot explain the out-of-plane spin alignment of the $\phi$ meson observed experimentally.
\end{abstract}

\maketitle

\section{Introduction}
The spin alignment, which quantifies the degree of anisotropy in the spin distribution of vector mesons, has come under intense scrutiny.
The state of a spin-1 particle is described by a $3\times3$ spin density matrix, $\rho_{\sigma\sigma'}$, with indices spanning $-1$, $0$, and $1$. The diagonal elements of the spin density matrix measure the probability of the particle being in the respective spin state. The components of the spin density matrix which determine the spin vector polarization, which is the measured one for the $\Lambda$ particle, are not measurable for vector mesons, due to the parity conserving decays they undergo. However, other components, and most notably the $\rho_{00}$ component, are measurable. If the particles are, on average, unpolarized, then all the diagonal components of the spin density matrix are  $1/3$. One observes a non-vanishing spin alignment if $\rho_{00}-1/3$ is non-zero. The spin alignment of vector mesons, in contrast with the spin polarization of Dirac fermions, is a spin tensor polarization phenomenon. We refer the reader to refs. \cite{Becattini:2024uha,Chen:2024afy} for recent reviews.

The STAR Collaboration has presented measurements of the global (i.e. integrated over all momentum space) and local (i.e. momentum-dependent) spin alignment of the $\phi$ and $K^{*}_0$ mesons \cite{STAR:2022fan} in AuAu collisions. On one hand, one sees that for the $\phi$ meson $\rho_{00}-1/3>0$, indicating a positive alignment especially for low collision energy. On the other, for $K^{*}_0$ one finds that the spin alignment is consistent with zero across all beam energy scans. This observation alone suggests that the spin alignment is strongly dependent on the vector meson species. This is different from the vector polarization of Dirac fermions, where the hydrodynamic mechanism driving it implies similar polarizations for particles of the same spin. An additional puzzle is given by the ALICE Collaboration measurements, where a negative spin alignment has been observed both for the $\phi$ and the $K^*_0$ at low transverse momentum, seemingly contradicting STAR's results \cite{ALICE:2019aid}. The ALICE Collaboration has also measured a negative alignmnet for the $J/\psi$ meson \cite{ALICE:2020iev,ALICE:2022dyy}.

The large $\phi$ spin alignment observed by STAR is particularly hard to explain theoretically. If a quark-coalescence model is used, and assuming that quarks are polarized by vorticity and shear, similarly to the $\Lambda$ hyperon, the ensuing alignment would be of the order of the $\Lambda$ polarization squared, $P_\Lambda^2\sim 10^{-4}$, which is two orders of magnitudes smaller than the measurement. On the statistical and hydrodynamic side, investigations have been carried considering higher order derivative contributions to alignment \cite{Zhang:2024mhs,Kumar:2023ojl}, dissipation \cite{Wagner:2022gza,Wagner:2023cct}, and the effect of non equilibrated spin degrees of freedom \cite{Goncalves:2024xzo,DeMoura:2023jzz}. Several alternative mechanisms have been proposed, including the effect of magnetic fields \cite{Yang:2017sdk,Sheng:2019kmk,Xia:2020tyd,Sheng:2022ssp,Li:2023tsf}, an intermediate fluctuating strong field \cite{Sheng:2019kmk,Sheng:2022wsy,Sheng:2023urn}, interaction with color fields \cite{Kumar:2022ylt,Muller:2021hpe,Kumar:2023ghs,Yang:2024qpy,Liang:2025hxw,Chen:2025mrf}, medium modifications to meson spectral functions \cite{Li:2022vmb,Grossi:2024pyh,Li:2024qae,Park:2022ayr,Sun:2025ror,Yin:2024dnu}, and even holographic models have been employed \cite{Zhao:2024ipr,Sheng:2024kgg}. Methods of open quantum systems have also being proposed \cite{Yang:2024ejk}. The most successful model to date is the one based on the fluctuations of the strong field. While these fluctuations would not contribute to the spin vector polarization of hyperons, averaging to zero, this model proposes that their correlations between constituent quark polarized by such field yield a the spin alignment of flavor singlet vector mesons \cite{Sheng:2019kmk,Sheng:2022wsy}. However, the first principle calculation of such correlations remains to be performed, and so far they are inferred from the data themselves.

In a previous paper \cite{Grossi:2024pyh}, we calculated the polarized emission rate of $\phi$ taking into account their interactions with a bath of Kaons in the hadronic phase, as well as dissipative corrections. Therefrom, we computed the spin density matrix in a static fluid and in a Bjorken flow, finding a very small effect. The data reported by STAR in $200$ GeV AuAu collision was consistent with our calculations. In this work we extend this approach to lower energy collisions, both taking into account the presence of nucleons in the hadronic phase and employing a realistic hydrodynamic simulation as a background. We tune our codes independently of spin alignment data, namely on $\phi$ transverse momentum dependent spectra, in order to have some sort of predictive power as concerning to the alignment itself. However, we find that our model doesn't yield a significant alignment, which remains consistent with zero ($\rho_{00}\simeq 1/3$) at variance with the data.

The paper is organized as follows. In section \ref{sec:rates} we summarize the formulae obtained in our previous work, as well as introduce the modification to the rates due to nucleons. Section \ref{sec:numerics} describes our numerical setup, as well as the tuning of the free parameters in our formulae, and section \ref{sec:results} shows the predictions we obtain concerning the spin alignment of the $\phi$ meson with our optimal parameters. We wrap up our findings in section \ref{sec:conclusions}, and report some supporting material in two appendices, \ref{app:scattering} and \ref{app:spectral}

\section{$\phi$ meson emission rate}\label{sec:rates}
In \cite{Grossi:2024pyh}, the emission rate of $\phi$ mesons interacting with Kaons was computed based on the chiral master formulae developed in \cite{Yamagishi:1995kr} for $SU(2)$ flavor symmetry and extended in
\cite{Kamano:2009st} for $U(3)$. Here we just recall our results, for future reference, and compute the contribution of nucleon scatterings to the rates.

The calculation starts from the equation:
\begin{equation}\label{eq:spin dep rate}
E_q\frac{\di R_{\sigma,\sigma'}}{\di^3q} = \frac{1}{e^{\beta q^0}+1}\frac{G_V^4 f_\phi^2}{(2\pi)^3 m_\phi^2}\epsilon^{\mu}_{\sigma}(q) {\epsilon^*}^{\nu}_{\sigma'}(q)   \text{Im}\left(i{W_{J_s}}^F_{\mu\nu}(q) \right),
\end{equation}
which connects the emission rate to the imaginary part of the Feynman propagator:
\begin{equation}
     W^F_{\mu\nu}(x) =\frac{1}{Z} \Tr \left[e^{-\beta H} T(J_{s\mu}(x) J_{s\nu}(0)) \right],
\end{equation}
where $J_s$ is the strange quark current. In the above equation, $f_\phi$ and $m_\phi$ are the decay constant and the mass of the $\phi$ meson, and $G_V$ is its coupling to strange currents.

The trace involved in the thermal propagator gets its leading contribution from the most densely populated states, so one can organize a virial expansion~\cite{Steele:1996su}:
\begin{align}
\label{PHINNX}
    W^F_{\mu\nu}(q)
    =&\int\di^4x e^{iq\cdot x}\langle 0|T ({J_s}_\mu(x){J_s}_\nu(0)) |0\rangle\nonumber\\
    &+\sum_a 
    \int \frac{\di^3k}{(2\pi)^3} \frac{n_K(k)}{2k^0}\int\di^4x e^{iqx}\langle K^a(k)|T (J_{s\mu}(x)J_{s\nu}(0) )|K^a(k)\rangle_{\text{conn.}}\nonumber\\
    &+\int \frac{\di^3p}{(2\pi)^3} \frac{n_N(p)}{2p^0}\int\di^4x e^{iqx}\sum_{spin}\langle N(p,s)|T (J_{s\mu}(x)J_{s\nu}(0) )|N(p,s)\rangle_{\text{conn.}}+\dots
\end{align}
The first two lines in the above expression were computed in \cite{Grossi:2024pyh}, and we have added the last term to account for the interaction of $\phi$ mesons with nucleons and resonances, which may become relevant in lower-energy collisions where the baryon density is non-negligible.

Within the vector dominance model, it is useful to rewrite the strange currents in terms of the $\phi$ field, using  $J_s^\mu=\frac{m_\phi^2}{G_V} \phi^\mu$. Then, the vacuum and the nucleon contribution are:
\begin{equation}
    \frac{m^4}{G_V^2}\left( \int\di^4x e^{iq\cdot x}\langle 0|T (\phi_\mu(x)\phi_\nu(0)) |0\rangle+\int \frac{\di^3p}{(2\pi)^3} \frac{n_N(p)}{2p^0}\sum_{spin}\langle N(p,s)|T (\phi_\mu(-q)\phi_\nu(q) )|N(p,s)\rangle_{\text{conn.}}\right).
\end{equation}
Note that both the Kaon and the nucleons contributions carry an integral over the density of states, which ultimately suppresses these corrections with respect to the vacuum one. Kaons are the most relevant state for this formalism because of their strangeness content, but their density is much lower compared to that of pions \cite{Grossi:2024pyh}. Pions would contribute to the $\phi$ emission rate if they had a strangeness content; although this has been suggested in~\cite{Braghin:2021hmr}, this effect is OZI suppressed.

The leading contribution to the emission rate of $\phi$ mesons in a resonance gas becomes:
 \begin{align}
 \label{eq:W0 res gas}
 E_q\frac{dR^0_{\sigma,\sigma'}}{d^3 q }  
 &=\delta_{\sigma,\sigma'} \bigg[\frac{1}{e^{\beta E^\phi_q}+1}m_\phi^2 G_V^2f^2_\phi   \frac{ q^2 }{ (2\pi)^3 }    \text{Im }\Pi_V^\phi(q)\bigg]_{q^2=m_\phi^2}.
 \end{align}
 One can see that this rate is spin-independent, and therefore no polarization can be produced.
The Kaon contribution is computed using the formalism of ref \cite{Yamagishi:1995kr,Kamano:2009st}, and one finds \cite{Grossi:2024pyh}:
\begin{align}\label{eq:rate1}
\frac{dR^K_{\sigma\sigma'}}{d^3 q }  =&\frac{1}{e^{\beta E^\phi_q}+1}\,{G_V^2}\, \frac{ 1  }{ (2\pi)^3 E^\phi_q } \,
\int \frac{d^3p}{(2\pi)^3}\frac 1{2E^K_p}\frac 1{e^{\beta E^K_p}-1}\nonumber\\
&\times\bigg[\frac {4q^2}{f_K^2}\, {\rm Im}\Pi_V^{88}(q)\delta_{\sigma\sigma'}
-((p\pm q)^2\delta_{\sigma\sigma'}-\epsilon_{\sigma}(q)\cdot p\, \epsilon^*_{\sigma'}(q)\cdot p)\frac 2{f_K^2}\,  {\rm Im}\Pi_A^{U}((p\pm q))\nonumber\\
&-q^2\epsilon_{\sigma}(q)\cdot p\epsilon^*_{\sigma'}(q)\cdot p\frac{8}{f_K^2}\,{\rm Re}\,\Delta_R(p\pm q)\, {\rm Im}\Pi_V^{88}(q)\bigg]_{q^2=m_\phi^2}.
\end{align}
On top of that, viscous corrections can be estimated. Using the approach of \cite{Liu:2017fib}, one finds:
\begin{align}
\label{RATEX}
\frac{dR^{\text{visc}}_{L,\perp}}{d^3 q }  =&\frac{1}{e^{\beta E^\phi_q}+1}\frac{G_V^4f^2_\phi}{m_\phi^2}  \frac{ 1  }{ (2\pi)^3 E^\phi_q } \,  \int \frac{d^3p}{(2\pi)^3}\frac 1{2E^K_p}\frac {e^{\beta E^K_p}}{(e^{\beta E^K_p}-1)^2}\frac {p^{\mu}p^\nu}{p\cdot u}
\left(\frac{t_\eta}{T}\sigma_{\mu\nu} +\frac{1}{3T}t_\zeta\theta\Delta_{\mu\nu}\right)\nonumber\\
&\times\bigg[\frac {4q^2}{f_K^2}\, {\rm Im}\Pi_V^{88}(q)
-((p\pm q)^2-|\epsilon_{L,\perp}\cdot (p\pm q)|^2)\frac 2{f_K^2}\,  {\rm Im}\Pi_A^{U}((p\pm q))\nonumber\\
&-q^2|\epsilon_{L,\perp}\cdot p|^2\frac{8}{f_K^2}\,{\rm Re}\,\Delta_R(p\pm q)\, {\rm Im}\Pi_V^{88}(q)\bigg]_{q^2=m_\phi^2},
 \end{align}
 where the expression has been written in covariant form. The shear tensor $\sigma$ and the expansion scalar $\theta$ read:
\begin{equation}
    \sigma_{\mu\nu}=\frac{1}{2}\left(\nabla_\mu u_\nu+\nabla_\nu u_\mu \right)-\frac{1}{3}\theta\Delta_{\mu\nu}\qquad \theta=\nabla_\mu u^\nu,
\end{equation}
with $\nabla_\mu =\Delta^{\nu}_{\mu}\partial_\nu$, $\Delta$ being the usual orthogonal projector to the foru-velocity $\Delta^{\mu\nu}=g^{\mu\nu}-u^\mu u^\nu$.
The shear $t_\eta$ and bulk $t_\zeta$ time scales are estimated as
\begin{align}
t_\eta\approx \frac{\eta}{e+P},\qquad \qquad t_\zeta\approx \frac{\zeta}{e+P},
\end{align}
where $\eta$ and $\zeta$ are the shear and bulk viscosity, respectively, and $e$ and $P$ are the thermodynamic energy density and pressure. In the Navier-Stokes limit, using $\pi^{\mu\nu}=2\eta\sigma_{\mu\nu}$ and $\Pi=-\zeta\theta$, one can also write the dissipative contribution as:
\begin{align}
\label{RATEX2}
\frac{dR^{\text{visc}}_{L,\perp}}{d^3 q }  =&\frac{1}{e^{\beta E^\phi_q}+1}\frac{G_V^4f^2_\phi}{m_\phi^2}  \frac{ 1  }{ (2\pi)^3 E^\phi_q } \,  \int \frac{d^3p}{(2\pi)^3}\frac 1{2E^K_p}\frac {e^{\beta E^K_p}}{(e^{\beta E^K_p}-1)^2}\frac {p^{\mu}p^\nu}{p\cdot u}
\left(\frac{\pi_{\mu\nu}}{2(e+P)T} -\frac{\theta \Delta_{\mu\nu}}{3(e+P)T}\right)\nonumber\\
&\times\bigg[\frac {4q^2}{f_K^2}\, {\rm Im}\Pi_V^{88}(q)
-((p\pm q)^2-|\epsilon_{L,\perp}\cdot (p\pm q)|^2)\frac 2{f_K^2}\,  {\rm Im}\Pi_A^{U}((p\pm q))\nonumber\\
&-q^2|\epsilon_{L,\perp}\cdot p|^2\frac{8}{f_K^2}\,{\rm Re}\,\Delta_R(p\pm q)\, {\rm Im}\Pi_V^{88}(q)\bigg]_{q^2=m_\phi^2},
 \end{align}
which will be the formula used numerically. We use the hadron resonance gas equation of state to calculate $e$ and $P$. 
Details about the spectral functions appearing in these formulae are reported briefly in appendix \ref{app:spectral}, and more details can be found in \cite{Grossi:2024pyh}.

We conclude this section by including the effects of nucleons. 
We assume a minimal interaction $\phi NN$ Lagrangian with Dirac and Pauli couplings:
\bea
\label{PHINN12}
{\cal L}_{\phi NN[\frac 12]}=
-g_{\phi N}\overline N\bigg(\phi_\mu\gamma^\mu-\frac{i\kappa_{\phi N}}{2m_N}\sigma^{\mu\nu}\partial_\mu\phi_\nu\bigg) N
\eea
with $N$ referring to the nucleon or its resonances, as the transition vertices of the type  nucleon-hyperon violates strangeness conservation. 
The contribution to eq. \eqref{PHINNX} comes from the $s$ and the $u$ channels of the $N\phi$ scatterings, which are calculated at tree level assuming effective values of the couplings. Details of the calculation are reported in appendix \ref{app:scattering}. Here we report the modification to the production rate, which reads:
\begin{align}
    E_q\frac{\di R^N_{\sigma\sigma'}}{\di^3q} = -
    \frac{1}{e^{\beta q^0}+1}\frac{G_V^2m_\phi^2 f_\phi^2}{(2\pi)^3}g_{\phi N}^2\left(f(q)\delta_{\sigma\sigma'}+g_{\sigma,\sigma'}(q)\right).
\end{align}  
Where the details of $f$ and $g$ are reported in appendix \ref{app:scattering}. We note that there is an isotropic contribution coming from $f$, and that all possible anisotropic polarization would come from $g$. Interestingly, $f$ is always non vanishing, both for stable nucleons and for resonances, whereas $g$ is non-vanishing for stable particles only in the $u$-channel, whereas for resonances both channels have $g$ terms, whose magnitude depends on the width of the resonance. In what follows, we have considered protons, neutrons, and the resonance $N(1880)$, with mass $m_N=1.88$GeV and width $\Gamma=0.3$ GeV \cite{pdg}. We have set $g_{\phi p}=g_{\phi n}=1$, $\kappa_{\phi p}=2.79$, and $\kappa_{\phi n}=-1.91$, the values of $\kappa$ corresponding to the magnetic moment of proton and neutron \cite{pdg}. As for the $N$ resonance, we have used $g_{\phi N}=-1.47$ and $\kappa_{\phi N}=-1.65$, fixed by the $\phi$-exchange contribution in the hyperon-nucleon-Nijmegen potential~\cite{PhysRevC.59.3009,PhysRevC.59.21}. The smallness of the couplings reflects on the OZI suppression rule.

\begin{figure}
    \centering
    \includegraphics[width=0.49\linewidth]{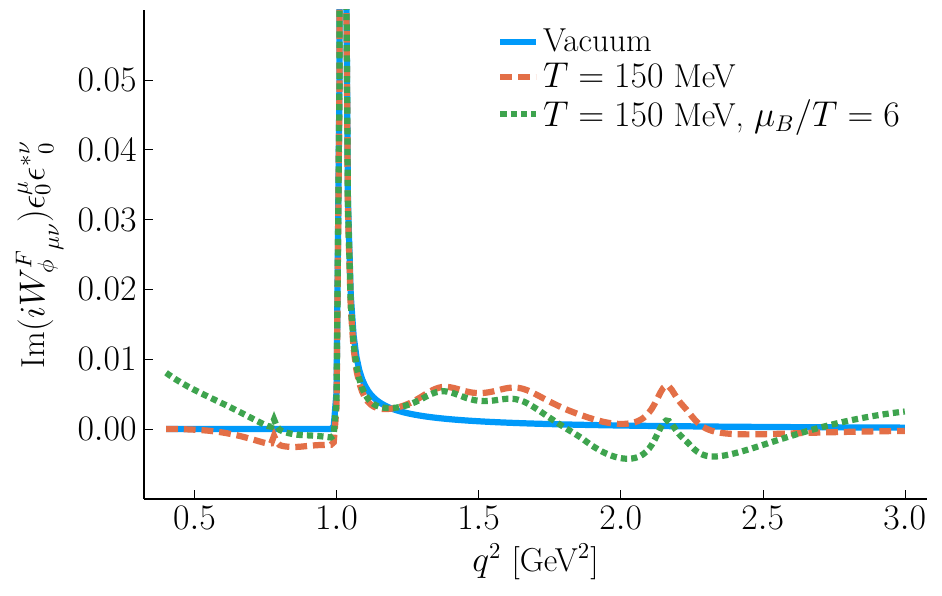}
    \caption{Modified vacuum ``$\phi$-spectral function'' by  matter effects}
    \label{fig:spectral}
\end{figure}

Figure \ref{fig:spectral} shows the matter modified vacuum ``$\phi$-spectral function'' entering 
the emission rates. The vacuum contribution  (solid-blue line), is modified by hadronic correlations at $T=150\,\rm MeV$ (dashed-orange line) mostly by  rescattering with kaons, and by a  baryon chemical potential $\mu_B/T=6$ (green-dotted line) by rescattering through nucleons. As stable hadronic matter is mostly dominated by pions, kaons and nucleons, strangeness is mostly lodged in kaons which are penalized at these
temperatures. The strangeness content of the pion and the nucleon is expected to be small, hence the small spectral deformations. To quantify these effects on $\phi$-production and correlations in heavy ion collisions, we now proceed to their numerical implementation using 
a realistic hydrodynamical  description of the collision process for different beam energies.

\section{Numerical analysis}\label{sec:numerics}

The total production rate for $\phi$-production in matter is given by
\begin{equation}
\label{eq:totalrate}
    \frac{\di R^{\textrm{tot.}}_{\sigma\sigma'}}{\di^3q}=
    \frac{\di R_{\sigma\sigma'}^0}{\di^3q}+\frac{\di R_{\sigma\sigma'}^K}{\di^3q}+\frac{\di R_{\sigma\sigma'}^{\textrm{visc.}}}{\di^3q}+\frac{\di R_{\sigma\sigma'}^N}{\di^3q}.
\end{equation}
from which the spin density matrix follows in the form
\begin{equation}\label{eq:spin density def}
    \rho_{\sigma\sigma'}(q)=\frac{\int \di^4 V\, E_q\frac{\di R^{\textrm{tot.}}_{\sigma\sigma'}}{\di^3q}}{\int \di^4 V\,\sum_\tau  E_q\frac{\di R^{\textrm{tot.}}_{\tau\tau}}{\di^3q}}
\end{equation}
where $\di^4V$ represents a space-time integral that takes place during the hadronic phase of the plasma. To evaluate eq.\eqref{eq:spin density def} one needs to calculate the rates in an evolving hadronic medium, and integrate over its spacetime history. In this work we use the code Fluid$u$m \cite{Floerchinger:2018pje,Devetak:2019lsk,Capellino:2023cxe}, to simulate the medium evolving according to viscous hydrodynamics.   

For our study, we have performed (2+1)D simulations (boost invariant fluid), with a constant shear and bulk viscosity over entropy density ratios. We have set $\eta/s=0.14$ and $\zeta/s=0.12$.
Charge diffusion is not implemented in Fluid$u$m, so we have modeled the chemical potential evolution as a Bjorken flow, such that at proper time $\tau=10$ fm the experimental estimates of freezout $\mu/T$ would be achieved. These values have been taken from \cite{STAR:2019bjj,Bollweg:2024epj}. We have run simulations at various collision energies, namely $\sqrt{s_{NN}}=\{$11.5, 19.6, 27, 39, 62.4, 200$\}$ GeV, corresponding to the energies explored in ref~\cite{STAR:2022fan}, where local and global alignment data for the $\phi$ meson is reported. For each of these energies, the baryon chemical potential to temperature ratio has been taken $\mu/T=\{$2.1, 1.3, 1.0, 0.7, 0.5, 0.2$\}$, respectively.

The initial state model is Trento \cite{Moreland:2014oya}. After sampling participants, we compute the thickness function and multiply it by a normalization factor, which is fixed later, to obtain the initial energy density. Having done that, we initialize the temperature from the Fluid$u$m equation of state. 

To constrain the free parameters of our model, in particular the coupling constant $G_V$ and the normalization of the initial state, we evaluate the transverse momentum dependent $\phi$ meson spectrum; this is nothing but the total number of $\phi$ mesons at a given $q_T$. Optimal parameters to reproduce experimental spectra, from refs. \cite{STAR:2008bgi,STAR:2015vvs}, have been fixed through a grid-search, minimizing the mean squared error between the model's prediction and the measurements. This procedure has been carried at all the collision energies of interest, taking into account all available centralities. For faster computations, and since the corrections due to interactions are suppressed by density, we have used the simple rate \eqref{eq:W0 res gas} for the grid search procedure. Our formulae hold in the hadronic phase, which has been identified with the fluid cells with a temperature $110$MeV$<T<170$MeV.

\begin{table}
    \centering
    \begin{tabular}{c|c|c|c|c|c|c}
        $\sqrt{s_{NN}}$ [GeV] & 11.5 & 19.6 & 27 & 39 & 62.4 & 200 \\
        \hline
        \hline
        $G_V$ & 1.3 & 1.6 & 1.5 & 1.5 & 2.0 & 2.0 \\
\end{tabular}
    \caption{Value of best-fit coupling constant at different collision energies.}
    \label{tab:params}
\end{table}

\begin{figure}
    \centering
    \includegraphics[width=0.49\linewidth]{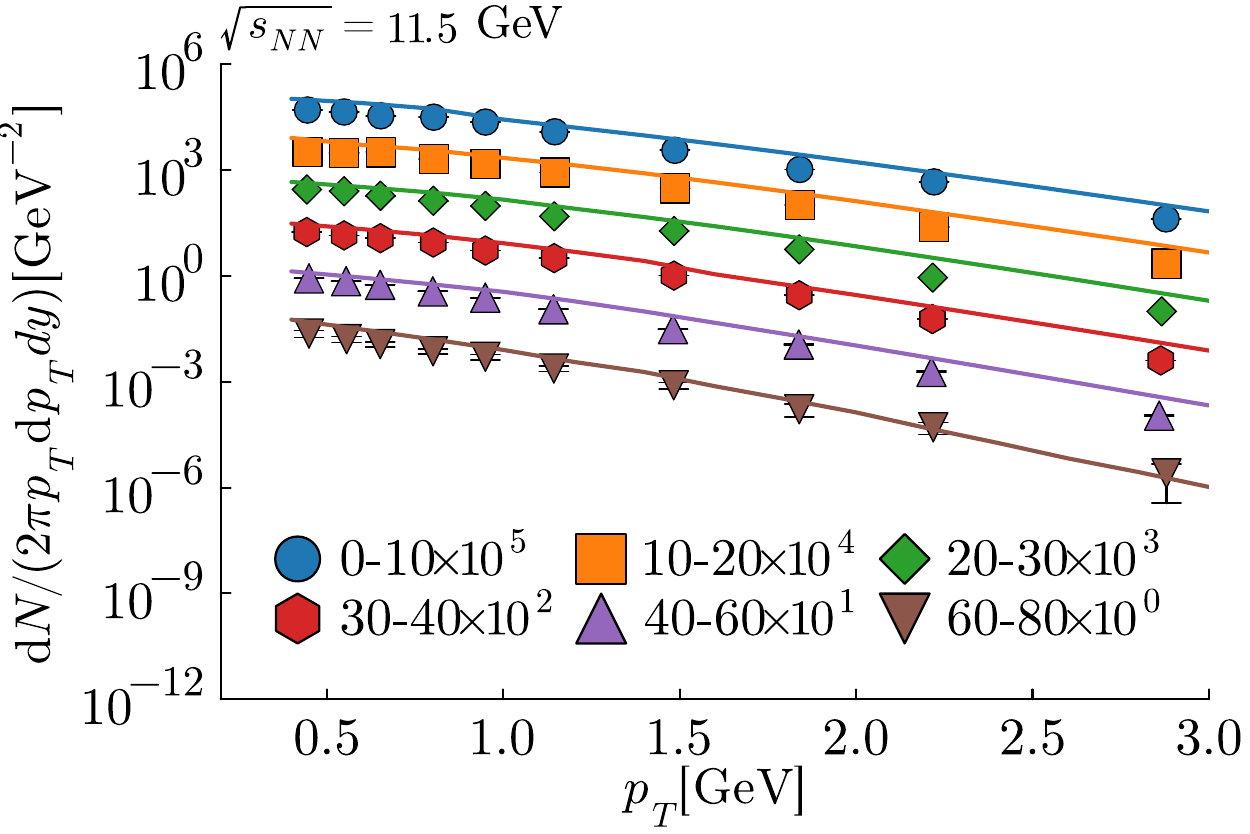}
    \includegraphics[width=0.49\linewidth]{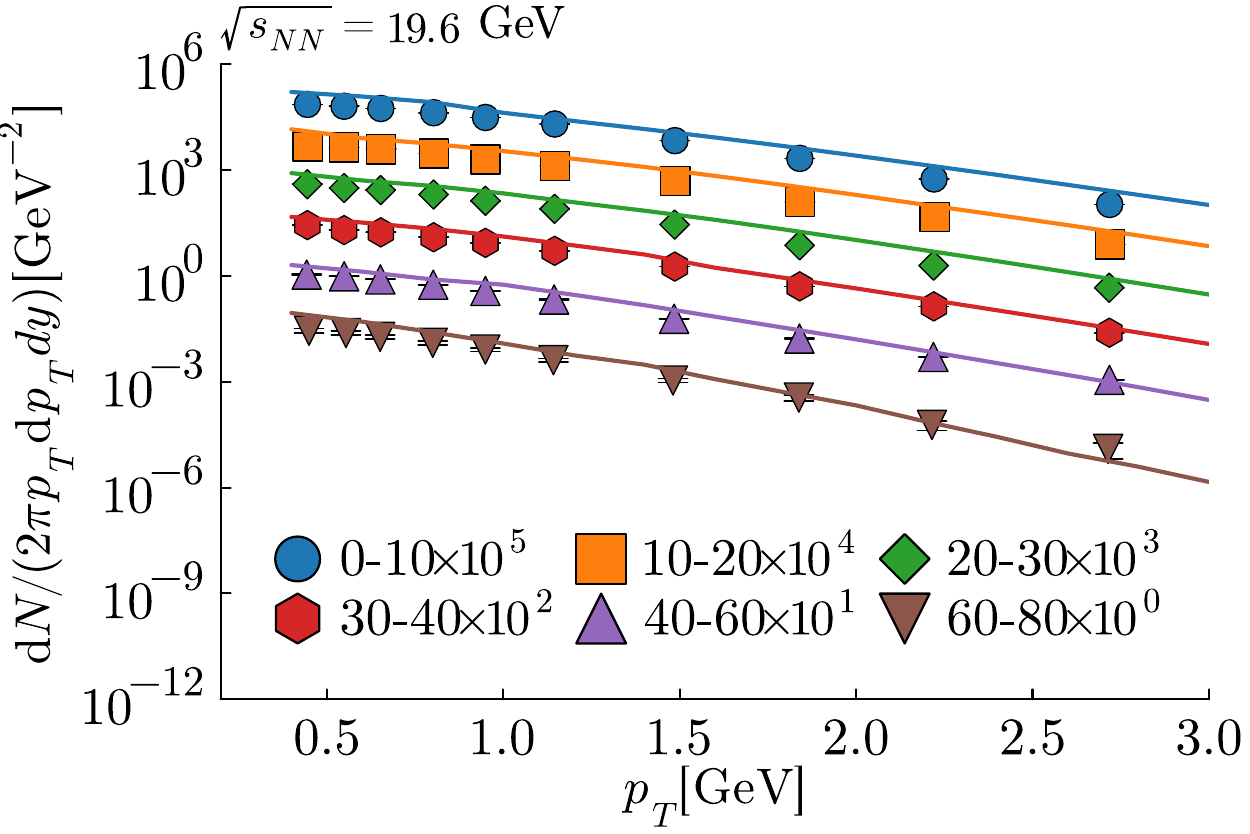}\\
    \includegraphics[width=0.49\linewidth]{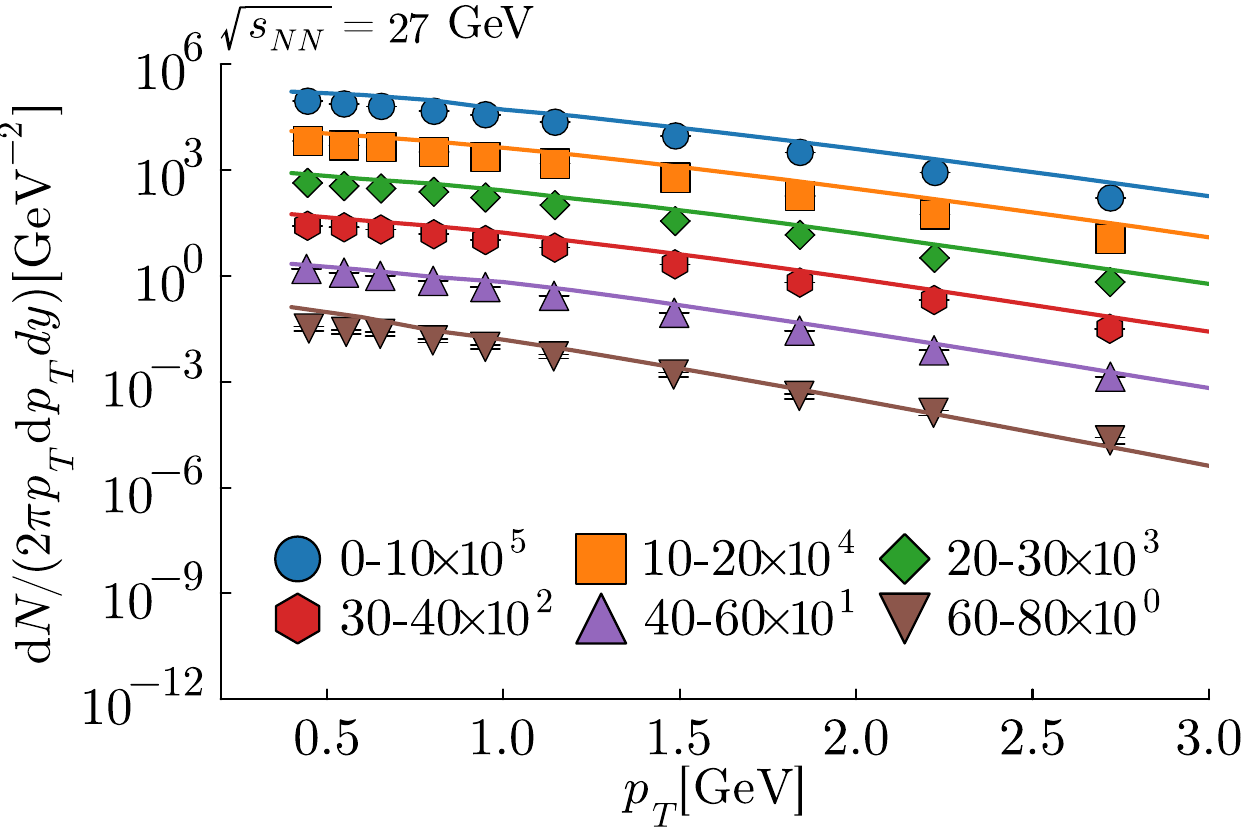}
    \includegraphics[width=0.49\linewidth]{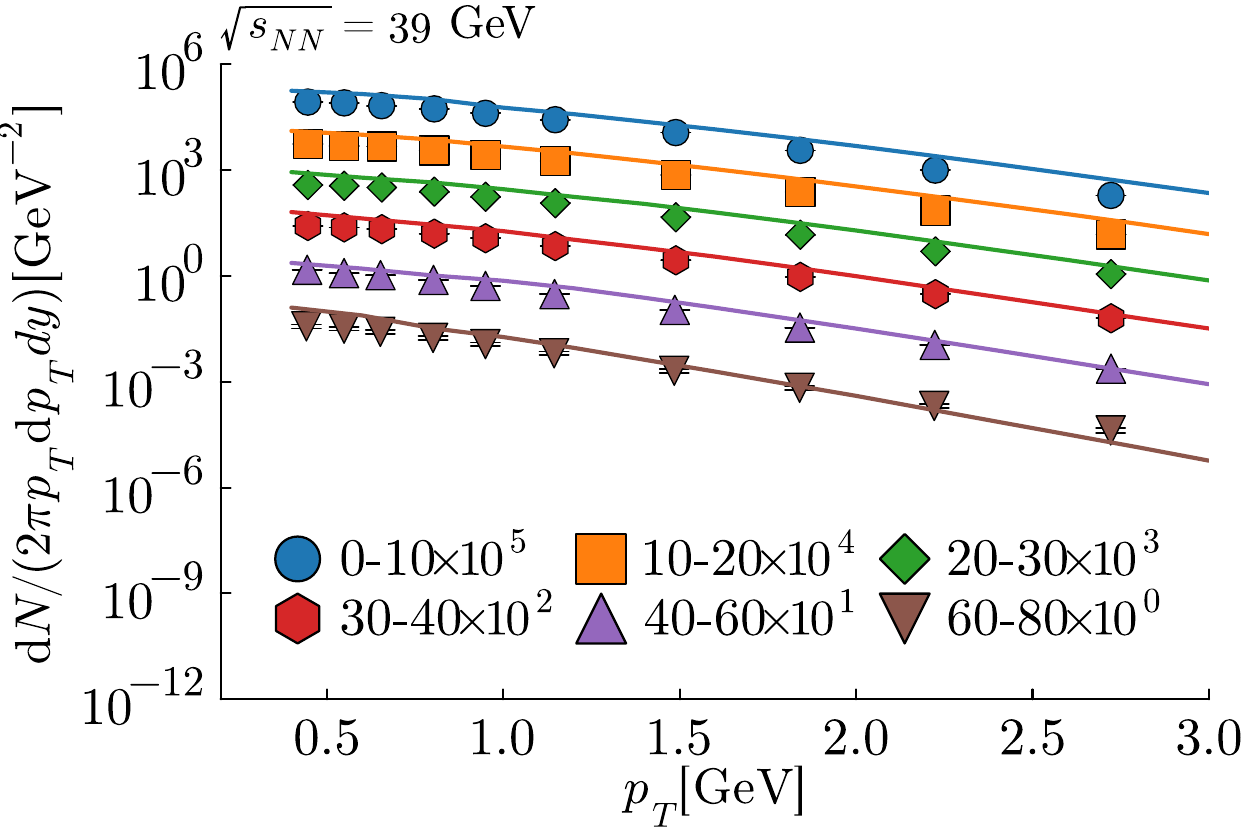}\\
    \includegraphics[width=0.49\linewidth]{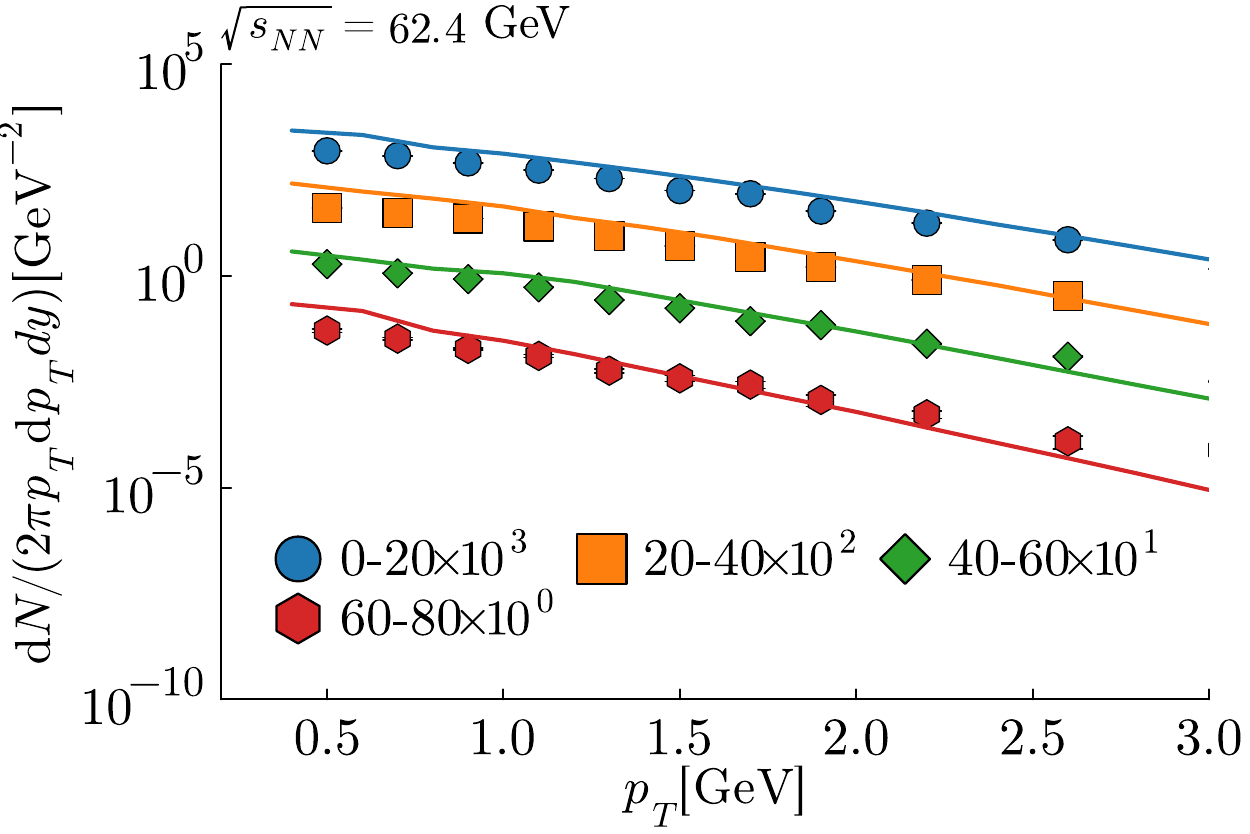}
    \includegraphics[width=0.49\linewidth]{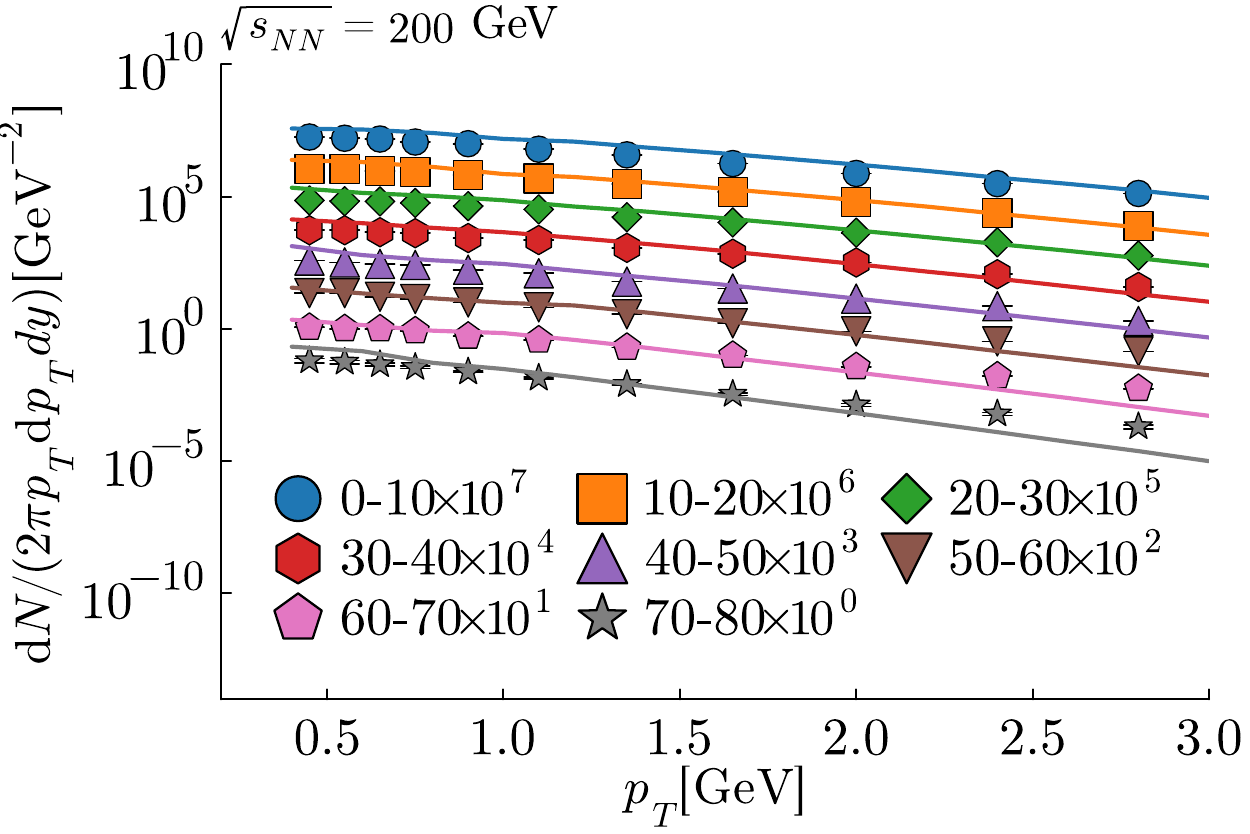}
    \caption{Transverse momentum dependent $\phi$ spectra at different centralities and center of mass energies $\sqrt{s_{NN}}$, which is reported in the title of each panel. Simulations are shown for the best fit values of table \ref{tab:params}. Data are taken from \cite{STAR:2008bgi,STAR:2015vvs}.}
    \label{fig:fit}
\end{figure}

The best fit values of $G_V$ are reported in table \ref{tab:params}, and the spectra are shown in figure \ref{fig:fit}, where the full rate eq. \eqref{eq:totalrate} is used. Notice that, to obtain a good fit of the spectra, the coupling appears to be a non-monotonic function of $\sqrt{s_{NN}}$. We obtain a satisfactory agreement between the model and the available data, especially for semi-peripheral collisions, which are in fact the most relevant to address the alignment data by STAR, which is taken in the centrality window $20$-$60\%$.  Now that the free parameters have been independently fixed, we can proceed to the calculation of alignment from the polarized emission rates.

\section{Results}\label{sec:results}
\begin{figure}
    \centering
    \includegraphics[width=0.49\linewidth]{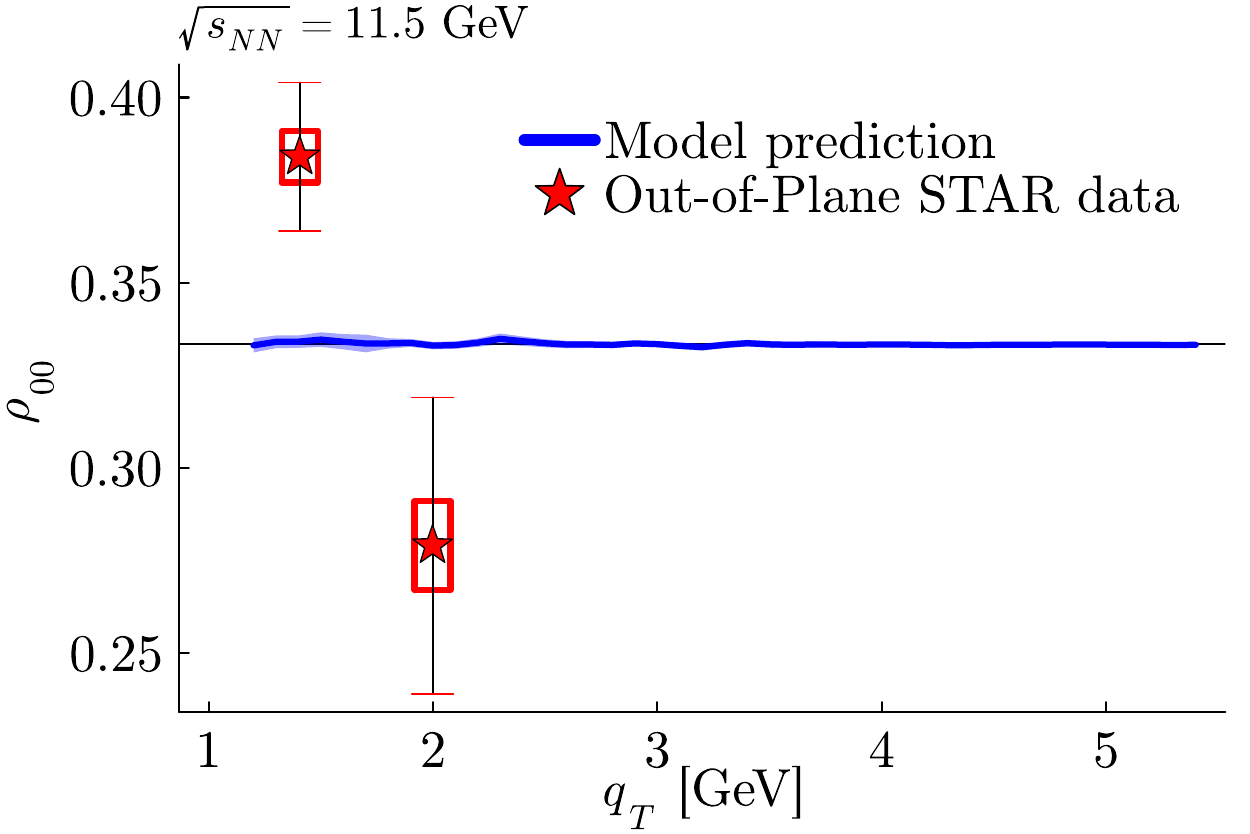}
    \includegraphics[width=0.49\linewidth]{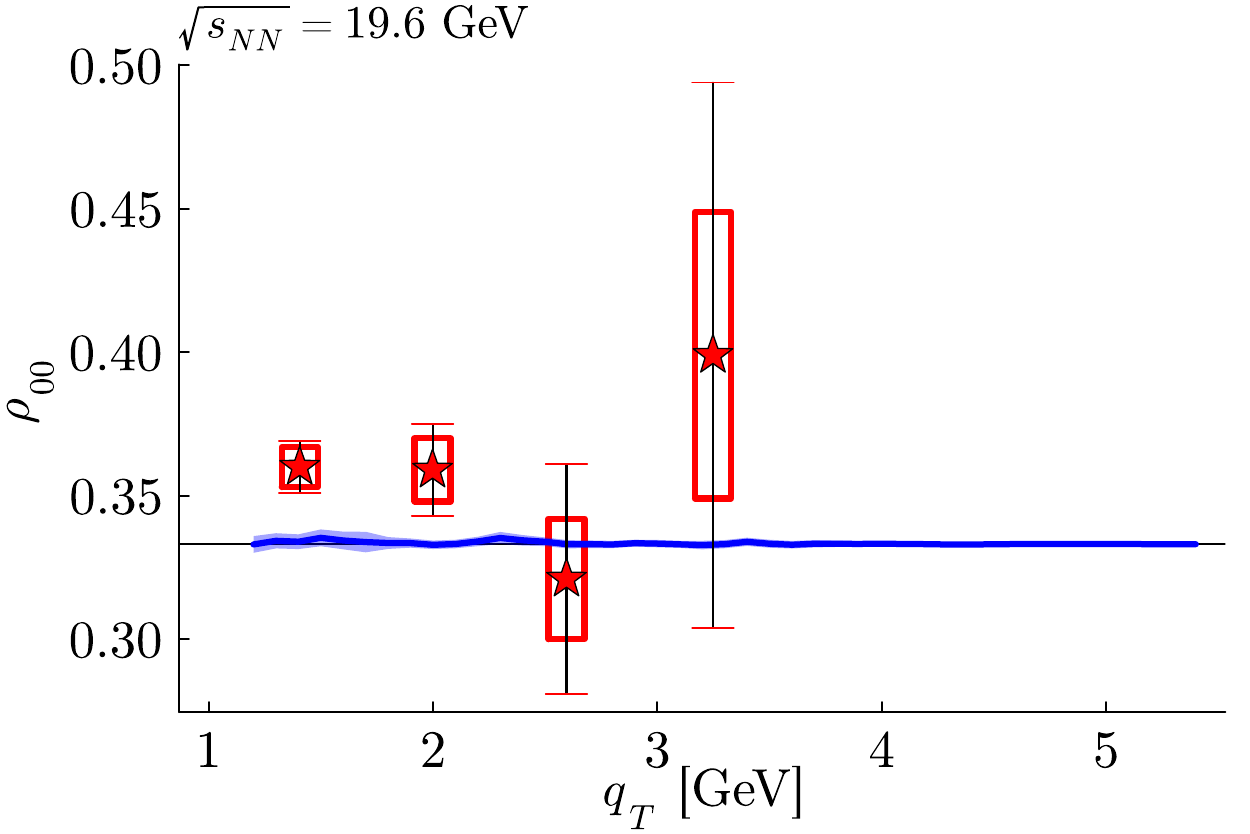}\\
    \includegraphics[width=0.49\linewidth]{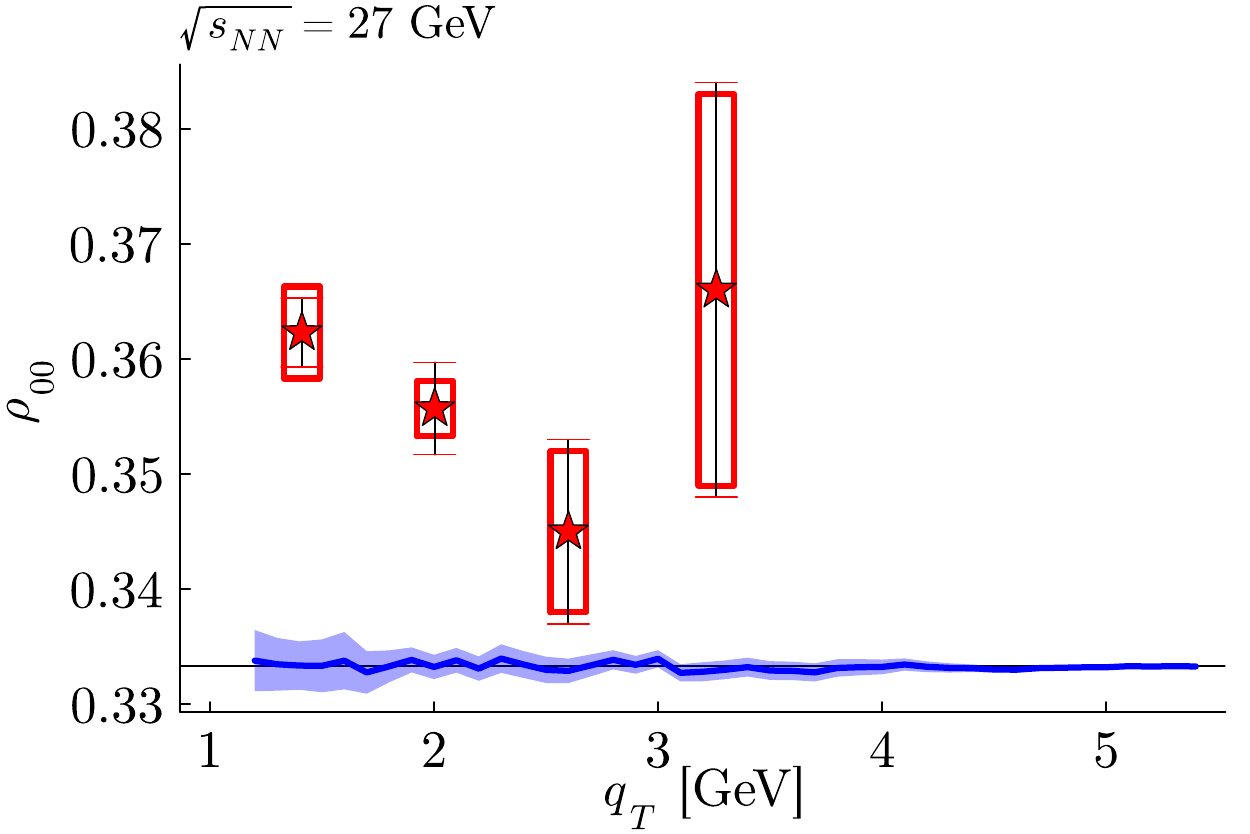}
    \includegraphics[width=0.49\linewidth]{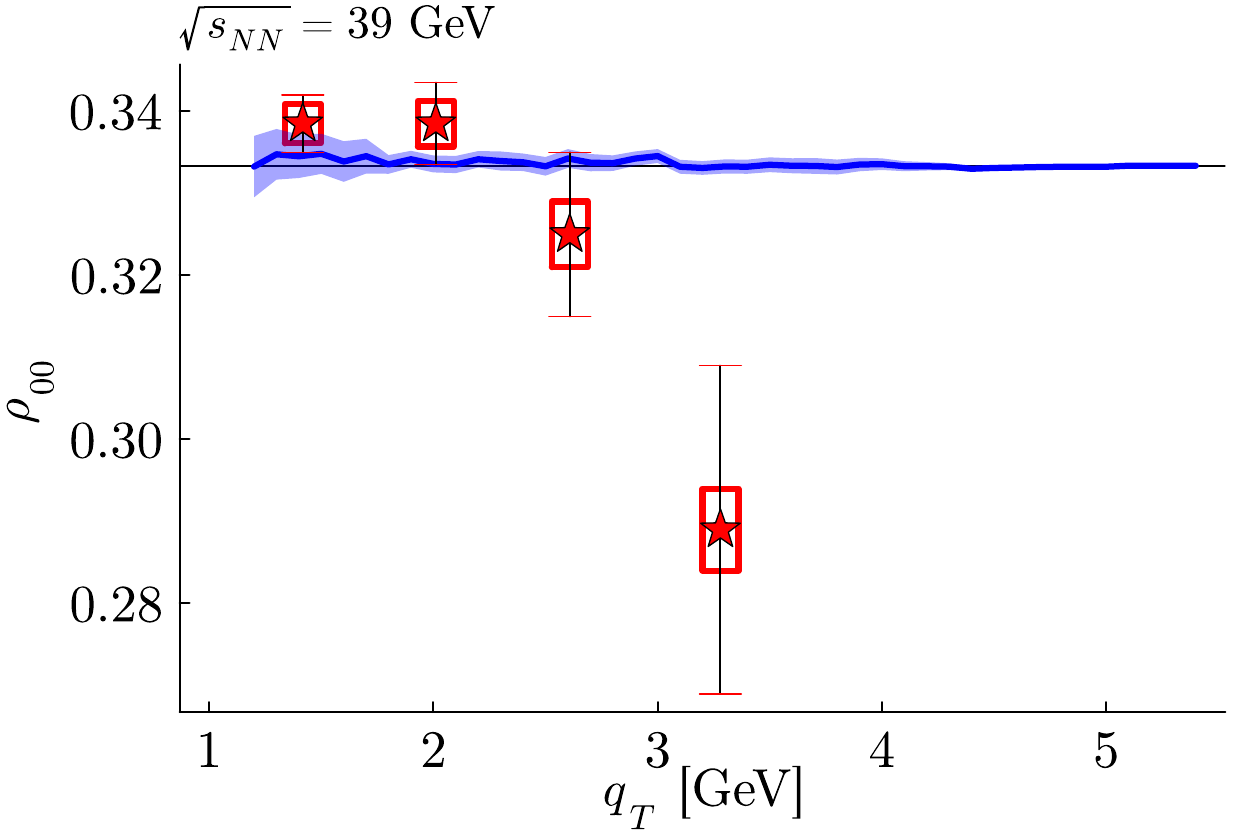}\\
    \includegraphics[width=0.49\linewidth]{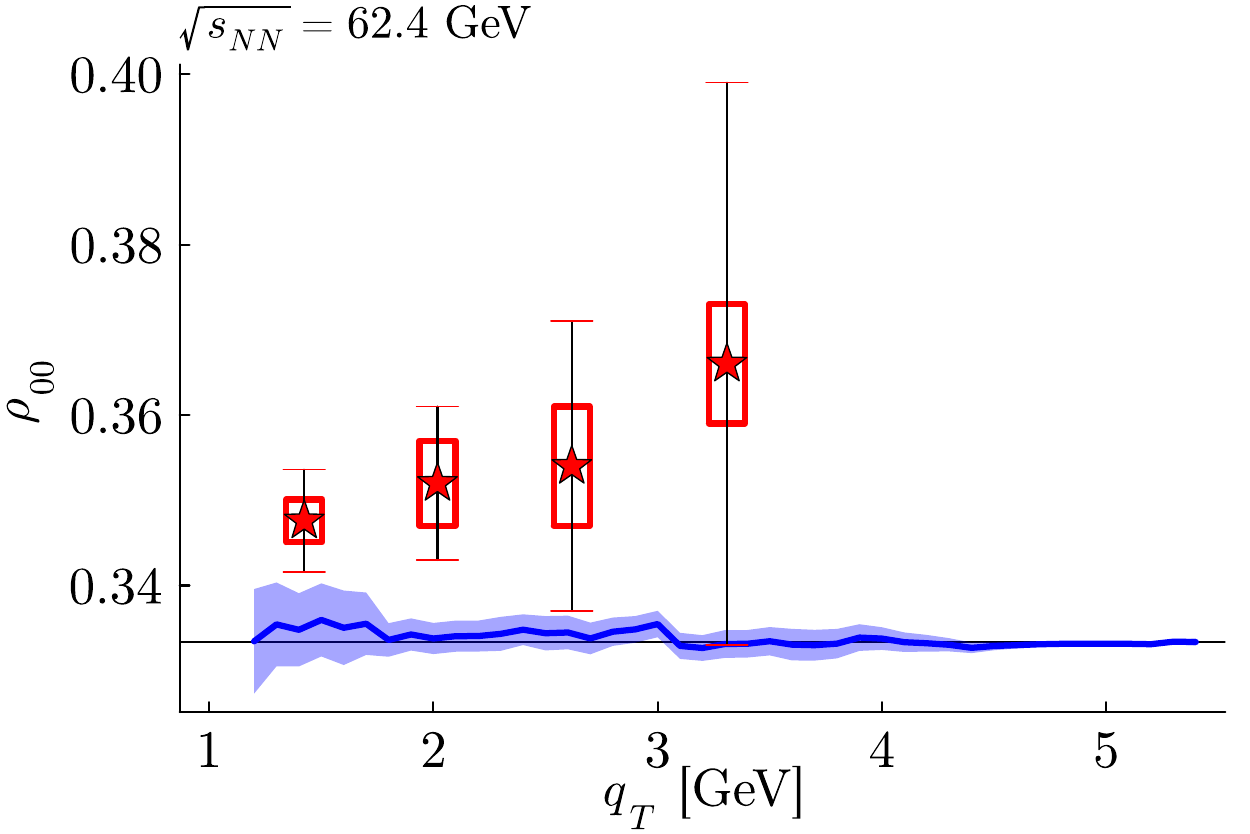}
    \includegraphics[width=0.49\linewidth]{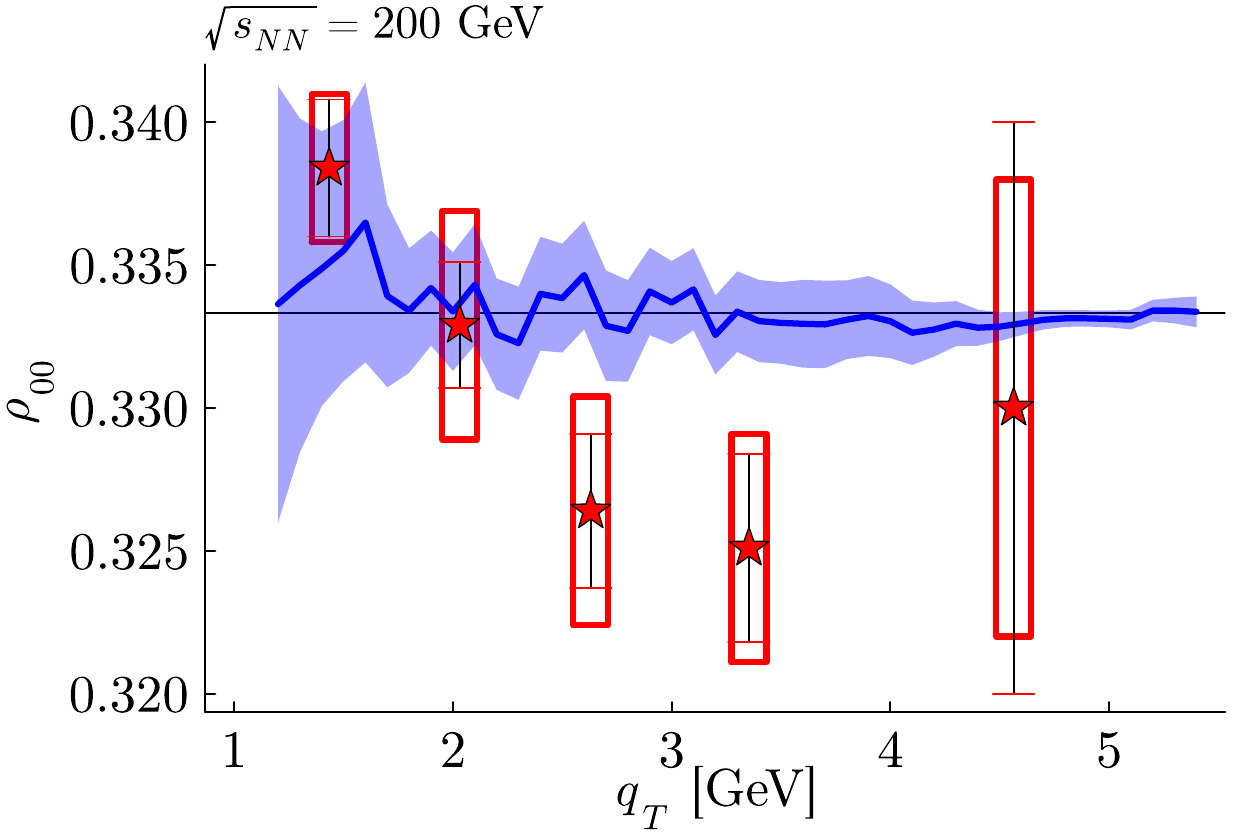}
    \caption{Local ($q_T$ dependent) spin alignment in the out of plane direction for different collision energies. Red stars represent STAR data \cite{STAR:2022fan}, and the blue line (plus statistical spread) is the model prediction. }
    \label{fig:local alignment}
\end{figure}

\begin{figure}
    \centering
    \includegraphics[width=0.49\linewidth]{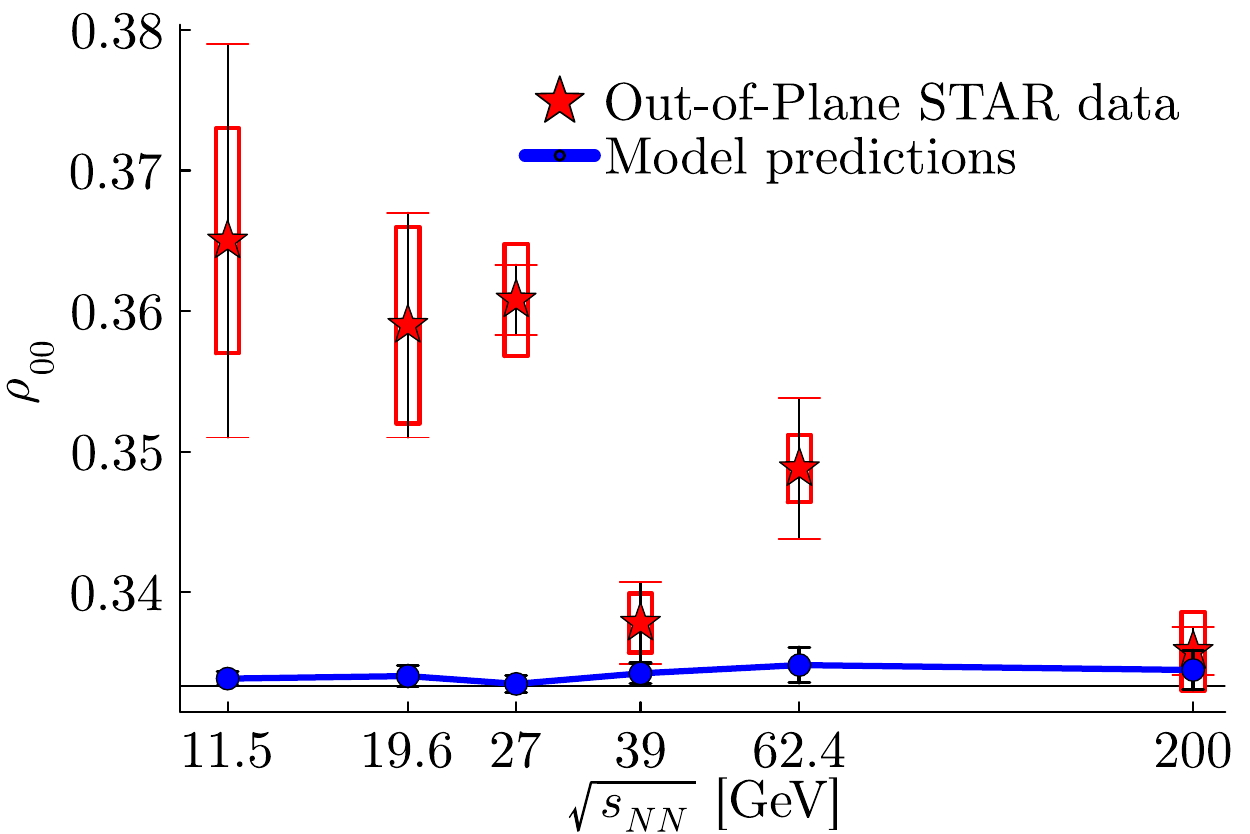}
    \includegraphics[width=0.49\linewidth]{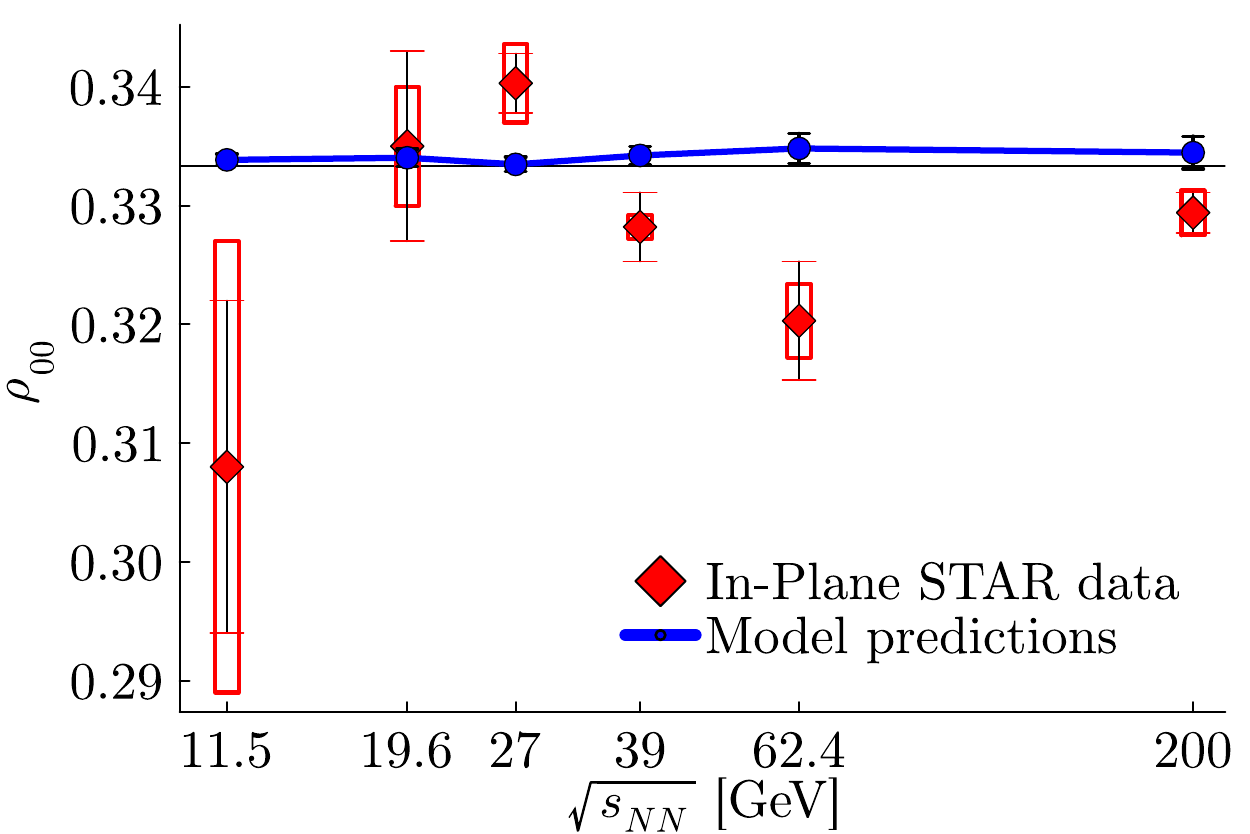}
    \caption{Collision energy dependence of the spin density matrix. Experimental data from ref \cite{STAR:2022fan} are compared to the predictions of our model. The right panel concerns out-of-plane alignment, where the quantization axis is taken along the $y$ direction, whereas the quantization axis is taken along the $z$ direction in the left hand panel, which shows in-plane alignment.}
    \label{fig:global alignment}
\end{figure}

We now proceed to evaluate the spin density matrix from eq. \eqref{eq:spin density def}. From these integrals we obtain both the transverse momentum-dependent and the $\sqrt{s_{NN}}$ dependent $\rho_{00}$. The results are reported in figures \ref{fig:local alignment} and \ref{fig:global alignment}, respectively.

Let us start discussing figure \ref{fig:local alignment} first. In this figure we show the transverse momentum dependence of $\rho_{00}$ for various collision energies. The data shown in these figures is taken from STAR \cite{STAR:2022fan}, and refers to the out-of-plane spin alignment. This means that the quantization axis of the $\phi$ meson is taken along the direction of the total angular momentum of the QGP. The simulations have been performed in the same centrality range, $20$-$60\%$, as the data, and the integration has been performed in the same kinematic window $|y|<1$, $y$ being the rapidity. Our model predicts that $\rho_{00}$ is consistent with $1/3$, at variance with experimental data.

From the local alignment, we can find the global one by integrating in $q_T$.
Namely, we define the global spin density matrix as:
\begin{equation}
    \rho_{\sigma\sigma'} =\frac{1}{N}\int\di N(q) \rho_{\sigma\sigma'}(q)= \frac{\int \frac{\di^3 q}{E_q} \left[ \int \di^4 V\, E_q \frac{\di R^{\textrm{tot.}}_{\sigma\sigma'}}{\di^3q} \right]}{\sum_\tau \int \frac{\di^3 q}{E_q} \left[ \int \di^4 V\, E_q \frac{\di R^{\textrm{tot.}}_{\tau\tau}}{\di^3q} \right]}, \qquad N=\sum_\tau \int \frac{\di^3 q}{E_q}  \int \di^4 V\, E_q \frac{\di R^{\textrm{tot.}}_{\tau\tau}}{\di^3q}
\end{equation}
where $N$ is the total number of produced $\phi$ according to our model.
In the right and left panels of figure \ref{fig:global alignment}, we compare the out-of-plane and in-plane STAR data to our model. The in-plane data assumes the direction of the quantization axis along the impact parameter of the collision.
We can see that the model predicts an isotropic distribution, with the alignment being at best very small. We also note that our model predicts the same alignment regardless the direction of the quantization axis. 

\begin{figure}
    \centering
    \includegraphics[width=0.49\linewidth]{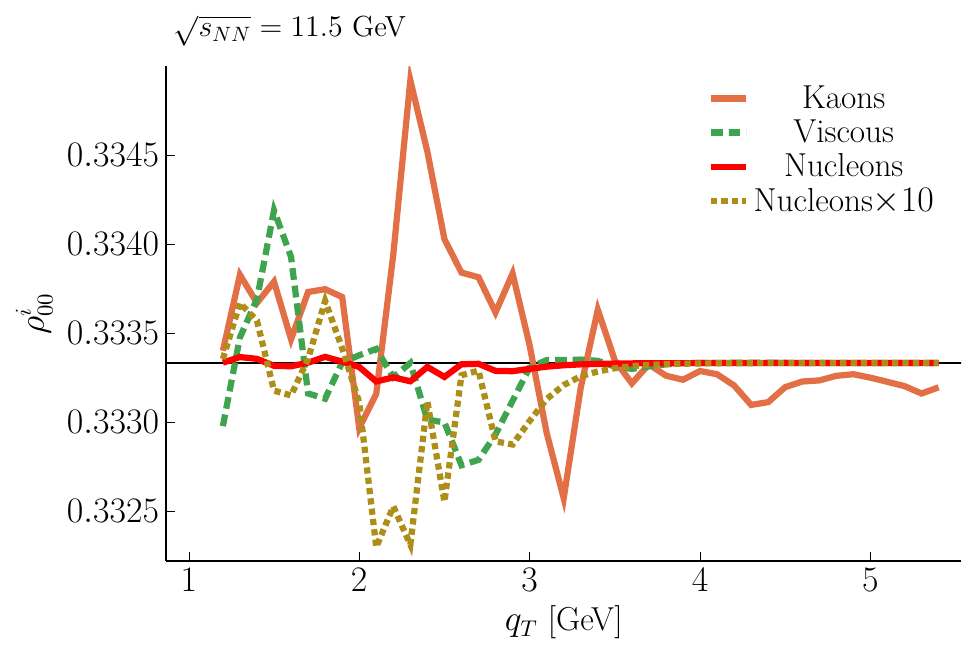}
    \caption{The different components contributing to spin alignment are shown individually. In the ``Nucleons$\times 10$'' entry, the the vacuum contribution to the alignment has not been modified, but $\di R^{\text{nucl}}/\di^3q$ in eq. \eqref{eq: rate component} has been multiplied by 10. Note that the $y$-axis scale has been changed from fig. \ref{fig:local alignment}, for clarity.}
    \label{fig:components}
\end{figure}

Finally, we report the result of individual contributions to alignment in figure \ref{fig:components}. This plot is obtained from the $\sqrt{s_{NN}}=11.5$ GeV simulation, where the baryon chemical potential is larger,  with expected  nucleon scattering effects. In this case, the individual contributions to the spin density matrix have been computed using 
\begin{equation}\label{eq: rate component}
    \frac{\di R_{\sigma\sigma'}}{\di^3q}=
    \frac{\di R_{\sigma\sigma'}^0}{\di^3q}+\frac{\di R_{\sigma\sigma'}^i}{\di^3q},
\end{equation}
in eq. \eqref{eq:spin density def}, $i$ selecting the Kaon, viscous, or nucleonic contributions. The dominant effect is caused by interactions with Kaons, with viscous corrections having a subleading importance. The effect of nucleon scattering appears to be substantially negligible, becoming comparable with the other two contributions only if enhanced by a factor 10, which would however imply unrealistically large couplings between the $\phi$ meson and baryons. We therefore conclude that nucleon scattering in the hadronic phase of the QGP do not play a significant role in determining the $\phi$ meson alignment,
over the range of the beam energy scan.

\section{Conclusion}\label{sec:conclusions}
In this paper, we have extended our previous work, ref. \cite{Grossi:2024pyh}, to a more realistic setup. We have employed a realistic hydrodynamic evolution to model the hadronic phase of the QGP, using a boost invariant setup. Free parameters of our theory have been fixed using available $\phi$ multiplicity data, and after that we have used our formalism to predict the ensuing spin alignment. Our result show that interactions with Kaons and nucleons in the hadronic phase are not enough to explain the large spin alignment observed experimentally. Notably, nucleon interactions seem to play a very little role, yielding a negligible contribution to alignment unless unrealistically large coupling constants are employed. We conclude that the origin of the $\phi$ meson spin alignment must be looked for in the earlier stages of the collision.

{\bf Acknowledgments} A.P. acknowledges useful discussions with F. Braghin and S. Liu at the 9th conference ``Chirality, vorticity, and magnetic fields in quantum matter''. The work of A.P. and I.Z. are supported by the Office of Science, U.S. Department of Energy under Contract No. DE-FG-88ER40388.

\appendix
\section{Scatterings of $\phi$ with nucleons}\label{app:scattering}
In this appendix we report the tree-level calculations for the scatterings between $\phi$ mesons and nucleons or resonances, denoted generically as a field $N$, with mass $M$ and, possibly, width $\Gamma$. Given the Lagrangian \eqref{PHINN12}, the s and u scattering channels at tree level correspond to the amplitudes:
\begin{align*}
    i{\cal A}^{\mu\nu}_s[\frac 12]=&
\frac{(-ig_{\phi N})^2}{s-m_N^2+i\Gamma  m_N}
\nonumber\\
&\times\bar{u}_s(p)\bigg(\gamma^\nu-\frac{\kappa_{\phi N}}{2m_N}\sigma^{\alpha\nu}q_\alpha\bigg)
i(\slashed{p}+\slashed{q}+m_{N})
\bigg(\gamma^\mu+\frac{\kappa_{\phi N}}{2m_N}\sigma^{\beta\mu}q_\beta\bigg)u_s(p)\\
i{\cal A}^{\mu\nu}_u[\frac 12]=&
\frac{(-ig_{\phi N})^2}{s-m_{N}^2+i\Gamma  m_{N}}
\nonumber\\
&\times\bar{u}_s(p)\bigg(\gamma^\mu+\frac{\kappa_{\phi N}}{2m_N}\sigma^{\alpha\mu}q_\alpha\bigg)
i(\slashed{q}-\slashed{p}+m_{N})
\bigg(\gamma^\nu-\frac{\kappa_{\phi N}}{2m_N}\sigma^{\beta\nu}q_\beta\bigg)u_s(p)
\end{align*}
After summing over final nucleon spin states, one gets:
\begin{align*}
    \Tr(i\mathcal{A}_s^{\mu\nu}) ={}&
    \frac{i g^2}{M^2 (M^2 - i \Gamma M - s)} \Bigg[
        (\kappa + 2i) M^2 \left( (\kappa + 2i) g^{\mu \nu} (p \cdot q)
        - 2i\, p^{\nu} q^{\mu} \right) \notag\\
        &\quad + p^{\mu} \Big(
            p^{\nu} \left( 8 M^2 - \kappa^2 (p \cdot q) \right)
            + 2 (2 - i\kappa) M^2 q^{\nu}
        \Big)
    \Bigg] \\
    \Tr(i\mathcal{A}_u^{\mu\nu}) ={}&
    -\frac{i g^2}{M^2 (M^2 - i \Gamma M - u)} \Bigg[
        (\kappa - 2i)^2 M^2 g^{\mu \nu} \left( 2 M^2 - p \cdot q \right) \notag\\
        &\quad + p^{\mu} \Big(
            p^{\nu} \left( \kappa^2 (p \cdot q) - 2 (\kappa - 2i)^2 M^2 \right)
            - 2i (\kappa - 2i) M^2 q^{\nu}
        \Big) \notag\\
        &\quad - 2i (\kappa - 2i) M^2 p^{\nu} q^{\mu}
    \Bigg]
\end{align*}
using $W_F=i \mathcal{A}$ and taking into account that terms with $q^{\mu,\nu}$ will drop once contracted with the polarization vectors in equation \eqref{eq:spin dep rate}, we have:
\begin{align*}
&\text{Im}\Tr(iW_s)=\text{Im}\left[i\Tr(i\mathcal{A})\right]=\nonumber\\
&\frac{g^2 \left[M {g}^{\mu \nu } \left({p}\cdot {q}\right) \left(-4 \kappa  M^2-\Gamma  \left(\kappa ^2-4\right) M+4 \kappa  s\right)+{p}^{\mu }{p}^{\nu } \left(\Gamma  \kappa ^2
   \left({p}\cdot {q}\right)-8 \Gamma  M^2\right)\right]}{M \left(M^2
   \left(\Gamma ^2-2 s\right)+M^4+s^2\right)}
   \\
   &\text{Im}\Tr(iW_u)=\\
&\frac{g^2 \left[M {g}^{\mu \nu } \left(4 \kappa (M^2-u)-\Gamma  \left(\kappa ^2-4\right) M\right) \left({p}\cdot
   {q}-2 M^2\right)+{p}^{\mu }{p}^{\nu } \left(\Gamma  \kappa^2 \left({p}\cdot {q}-2M^2\right)+ 8M  \left(\Gamma M-\kappa(u-M^2)\right)\right)\right]}{M \left(M^2 \left(\Gamma ^2-2 u\right)+M^4+u^2\right)}
\end{align*}
now, taking into account $\epsilon_s\cdot \epsilon^*_r=-\delta_{rs}$, the rate can be written as:
    \begin{align}
    E_q\frac{\di R_{\sigma,\sigma'}}{\di^3q} =-    \frac{1}{e^{\beta q^0}+1}\frac{G_V^2m_\phi^2 f_\phi^2}{(2\pi)^3}g^2\left(f^{(s,u)}(q)\delta_{\sigma\sigma'}+g^{(s,u)}_{\sigma,\sigma'}(q)\right).
    \end{align}    
    with
    \begin{align}
        &f^{(s)}(q)= 
        \int\frac{\di^3p}{(2\pi)^3}\frac{n_N(p)}{2p^0} 
    \; p\cdot q \;\frac{ 4\kappa(s-M^2)-\Gamma M(\kappa^2-4)}{(s-M^2)^2+M^2\Gamma^2} ,\\
      &g^{(s)}_{\sigma,\sigma'}(q)=-\frac{\Gamma }{M}
      \int\frac{\di^3p}{(2\pi)^3}\frac{n_N(p)}{2p^0}\;p\cdot \epsilon_\sigma p\cdot \epsilon_{\sigma'}\;\frac{ \kappa^2\;p\cdot q -8 M^2}{(s-M^2)^2+M^2\Gamma^2}\\
      &f^{(u)}(q)=   \int\frac{\di^3p}{(2\pi)^3}\frac{n_N(p)}{2p^0} 
    \left( p\cdot q -2M^2\right)\frac{ 4\kappa(M^2-u)-\Gamma M(\kappa^2-4)}{(u-M^2)^2+M^2\Gamma^2} ,\\
      &g^{(u)}_{\sigma,\sigma'}(q)=-
      \int\frac{\di^3p}{(2\pi)^3}\frac{n_N(p)}{2p^0}\;p\cdot \epsilon_\sigma p\cdot \epsilon_{\sigma'}\;\frac{ \Gamma\kappa^2(\frac{p\cdot q}{M}-2M)+8(\Gamma M+\kappa(M^2-u))}{(u-M^2)^2+M^2\Gamma^2}
\end{align}

\section{Spectral functions}\label{app:spectral}
We report, for the reader's convenience, the details of the spectral functions used in this and in our previous work \cite{Grossi:2024pyh}.
For a particle with mass $m$, the vector and axial SU(2) iso-spectral functions are:
\begin{align}
\label{PIVA}
\Pi_V^I(q^2)=\frac {f_V^2}{q^2}({\bf F}_V(q^2)-1), \qquad 
\Pi_A^I(q^2)=\frac{f_V^2}{m_V^2-q^2-im_V\Gamma(q^2)},
\end{align}
with:
\begin{align}
{\bf F}_V(q^2)=&\frac{m_V^2}{m_V^2-q^2-im_V \Gamma(q^2)}\nonumber\\
\Gamma(q^2)=&\theta(q^2-M^2_{BR})\Gamma_{0}\frac{m_V}{\sqrt{q^2}}\bigg(\frac{q^2-M^2_{BR}}{m_V^2-M^2_{BR}}\bigg)^{\frac 32}
\end{align}
$\theta (x)$ being the step function. The $M^2_{BR}$ value depends on the particle under consideration and its main decay channel.  For the lightest isoscalar resonances, one can use the standard decomposition
\begin{align}
\omega_8=\sqrt{\frac 13}\omega -\sqrt{\frac 23}\phi\qquad\rightarrow\qquad \Pi_V^{88}=\frac 13 \Pi_V^\omega+\frac 23 \Pi_V^\phi.
\end{align}
The U-spin axial spectral function is
\begin{align}
   \Pi_A^{U}=\Pi_A^{K_1(1270)} +\Pi_A^{K_1(1400)} 
\end{align}
and involves the $\frac 12 (1^+)$ resonances  $K_1(1270)$ and $K_1(1400)$. Additional resonances are also considered in the appropriate spectral function, depending on their quantum numbers \cite{pdg}. The characteristics of the particles considered are reported in table \ref{tab:particles}.
\begin{table}
    \centering
    \begin{tabular}{c c c c c c}
    \hline
    \hline
     & $I(J^{PC})$ & $m$[MeV]   &  $f$[MeV] & $\Gamma_0$[MeV] & $M_{BR}$\\
    \hline
    \hline
    $\omega$ & $0(1^{--})$ & 782.66 & $44.26$ & 8.68 & $3m_\pi$\\
    $\phi$ & $0(1^{--})$ & 1019.461 & $-81.53$ & 4.249 & $2m_K$\\
    $\omega(1420)$ & $0(1^{--})$ & 1410 & $290$ & 290 & $m_\rho+m_\pi$\\
    $\phi(1680)$ & $0(1^{--})$ & 1680 & $150$ & 150 & $2m_K+m_\pi$\\
    $\omega(1650)$ & $0(1^{--})$ & 1670 & $315$ & 315 & $m_\pi+m_\rho$\\
    $\phi(2170)$ & $0(1^{--})$ & 2164 & $106$ & 106 & $m_\phi+2m_\pi$\\
    $\omega(2220)$ & $0(1^{--})$ & 2232 & $93$ & 93 & $2m_\pi+m_\omega$\\
    \hline
    $K_1$(1270) & $\frac{1}{2}(1^{+})$ &1253 & 90& 90& $m_\rho+m_K$ \\
    $K_1$(1400) & $\frac{1}{2}(1^{+})$ &1403 & 174& 174& $m_{K^*(892)}+m_\pi$ \\
    $K_1$(1670) & $\frac{1}{2}(1^{+})$ &1650 & 150& 150& $m_{K}+2m_\pi$ \\
    \hline
    \hline
    \end{tabular}
    \caption{The properties of particles appearing in the relevant spectral functions. The decay constants of $\omega$ and $\phi$ are obtained using the KFSR relations, whereas for the other resonances, we have used $f=\Gamma_0$. Masses and decay widths are taken from ref.~\cite{pdg}.}
    \label{tab:particles}
\end{table}
For further details we refer the reader to \cite{Grossi:2024pyh}.

\bibliography{biblio}

\begin{thebibliography}{50}%
\makeatletter
\providecommand \@ifxundefined [1]{%
 \@ifx{#1\undefined}
}%
\providecommand \@ifnum [1]{%
 \ifnum #1\expandafter \@firstoftwo
 \else \expandafter \@secondoftwo
 \fi
}%
\providecommand \@ifx [1]{%
 \ifx #1\expandafter \@firstoftwo
 \else \expandafter \@secondoftwo
 \fi
}%
\providecommand \natexlab [1]{#1}%
\providecommand \enquote  [1]{``#1''}%
\providecommand \bibnamefont  [1]{#1}%
\providecommand \bibfnamefont [1]{#1}%
\providecommand \citenamefont [1]{#1}%
\providecommand \href@noop [0]{\@secondoftwo}%
\providecommand \href [0]{\begingroup \@sanitize@url \@href}%
\providecommand \@href[1]{\@@startlink{#1}\@@href}%
\providecommand \@@href[1]{\endgroup#1\@@endlink}%
\providecommand \@sanitize@url [0]{\catcode `\\12\catcode `\$12\catcode
  `\&12\catcode `\#12\catcode `\^12\catcode `\_12\catcode `\%12\relax}%
\providecommand \@@startlink[1]{}%
\providecommand \@@endlink[0]{}%
\providecommand \url  [0]{\begingroup\@sanitize@url \@url }%
\providecommand \@url [1]{\endgroup\@href {#1}{\urlprefix }}%
\providecommand \urlprefix  [0]{URL }%
\providecommand \Eprint [0]{\href }%
\providecommand \doibase [0]{http://dx.doi.org/}%
\providecommand \selectlanguage [0]{\@gobble}%
\providecommand \bibinfo  [0]{\@secondoftwo}%
\providecommand \bibfield  [0]{\@secondoftwo}%
\providecommand \translation [1]{[#1]}%
\providecommand \BibitemOpen [0]{}%
\providecommand \bibitemStop [0]{}%
\providecommand \bibitemNoStop [0]{.\EOS\space}%
\providecommand \EOS [0]{\spacefactor3000\relax}%
\providecommand \BibitemShut  [1]{\csname bibitem#1\endcsname}%
\let\auto@bib@innerbib\@empty
\bibitem [{\citenamefont {Becattini}\ \emph {et~al.}(2024)\citenamefont
  {Becattini}, \citenamefont {Buzzegoli}, \citenamefont {Niida}, \citenamefont
  {Pu}, \citenamefont {Tang},\ and\ \citenamefont {Wang}}]{Becattini:2024uha}%
  \BibitemOpen
  \bibfield  {author} {\bibinfo {author} {\bibfnamefont {F.}~\bibnamefont
  {Becattini}}, \bibinfo {author} {\bibfnamefont {M.}~\bibnamefont
  {Buzzegoli}}, \bibinfo {author} {\bibfnamefont {T.}~\bibnamefont {Niida}},
  \bibinfo {author} {\bibfnamefont {S.}~\bibnamefont {Pu}}, \bibinfo {author}
  {\bibfnamefont {A.-H.}\ \bibnamefont {Tang}}, \ and\ \bibinfo {author}
  {\bibfnamefont {Q.}~\bibnamefont {Wang}},\ }\href@noop {} {\  (\bibinfo
  {year} {2024})},\ \Eprint {http://arxiv.org/abs/2402.04540} {arXiv:2402.04540
  [nucl-th]} \BibitemShut {NoStop}%
\bibitem [{\citenamefont {Chen}\ \emph {et~al.}(2025)\citenamefont {Chen},
  \citenamefont {Liang}, \citenamefont {Ma}, \citenamefont {Sheng},\ and\
  \citenamefont {Wang}}]{Chen:2024afy}%
  \BibitemOpen
  \bibfield  {author} {\bibinfo {author} {\bibfnamefont {J.-H.}\ \bibnamefont
  {Chen}}, \bibinfo {author} {\bibfnamefont {Z.-T.}\ \bibnamefont {Liang}},
  \bibinfo {author} {\bibfnamefont {Y.-G.}\ \bibnamefont {Ma}}, \bibinfo
  {author} {\bibfnamefont {X.-L.}\ \bibnamefont {Sheng}}, \ and\ \bibinfo
  {author} {\bibfnamefont {Q.}~\bibnamefont {Wang}},\ }\href {\doibase
  10.1007/s11433-024-2495-1} {\bibfield  {journal} {\bibinfo  {journal} {Sci.
  China Phys. Mech. Astron.}\ }\textbf {\bibinfo {volume} {68}},\ \bibinfo
  {pages} {211001} (\bibinfo {year} {2025})},\ \Eprint
  {http://arxiv.org/abs/2407.06480} {arXiv:2407.06480 [hep-ph]} \BibitemShut
  {NoStop}%
\bibitem [{\citenamefont {Abdallah}\ \emph {et~al.}(2023)\citenamefont
  {Abdallah} \emph {et~al.}}]{STAR:2022fan}%
  \BibitemOpen
  \bibfield  {author} {\bibinfo {author} {\bibfnamefont {M.~S.}\ \bibnamefont
  {Abdallah}} \emph {et~al.} (\bibinfo {collaboration} {STAR}),\ }\href
  {\doibase 10.1038/s41586-022-05557-5} {\bibfield  {journal} {\bibinfo
  {journal} {Nature}\ }\textbf {\bibinfo {volume} {614}},\ \bibinfo {pages}
  {244} (\bibinfo {year} {2023})},\ \Eprint {http://arxiv.org/abs/2204.02302}
  {arXiv:2204.02302 [hep-ph]} \BibitemShut {NoStop}%
\bibitem [{\citenamefont {Acharya}\ \emph {et~al.}(2020)\citenamefont {Acharya}
  \emph {et~al.}}]{ALICE:2019aid}%
  \BibitemOpen
  \bibfield  {author} {\bibinfo {author} {\bibfnamefont {S.}~\bibnamefont
  {Acharya}} \emph {et~al.} (\bibinfo {collaboration} {ALICE}),\ }\href
  {\doibase 10.1103/PhysRevLett.125.012301} {\bibfield  {journal} {\bibinfo
  {journal} {Phys. Rev. Lett.}\ }\textbf {\bibinfo {volume} {125}},\ \bibinfo
  {pages} {012301} (\bibinfo {year} {2020})},\ \Eprint
  {http://arxiv.org/abs/1910.14408} {arXiv:1910.14408 [nucl-ex]} \BibitemShut
  {NoStop}%
\bibitem [{\citenamefont {Acharya}\ \emph {et~al.}(2021)\citenamefont {Acharya}
  \emph {et~al.}}]{ALICE:2020iev}%
  \BibitemOpen
  \bibfield  {author} {\bibinfo {author} {\bibfnamefont {S.}~\bibnamefont
  {Acharya}} \emph {et~al.} (\bibinfo {collaboration} {ALICE}),\ }\href
  {\doibase 10.1016/j.physletb.2021.136146} {\bibfield  {journal} {\bibinfo
  {journal} {Phys. Lett. B}\ }\textbf {\bibinfo {volume} {815}},\ \bibinfo
  {pages} {136146} (\bibinfo {year} {2021})},\ \Eprint
  {http://arxiv.org/abs/2005.11128} {arXiv:2005.11128 [nucl-ex]} \BibitemShut
  {NoStop}%
\bibitem [{\citenamefont {Acharya}\ \emph {et~al.}(2023)\citenamefont {Acharya}
  \emph {et~al.}}]{ALICE:2022dyy}%
  \BibitemOpen
  \bibfield  {author} {\bibinfo {author} {\bibfnamefont {S.}~\bibnamefont
  {Acharya}} \emph {et~al.} (\bibinfo {collaboration} {ALICE}),\ }\href
  {\doibase 10.1103/PhysRevLett.131.042303} {\bibfield  {journal} {\bibinfo
  {journal} {Phys. Rev. Lett.}\ }\textbf {\bibinfo {volume} {131}},\ \bibinfo
  {pages} {042303} (\bibinfo {year} {2023})},\ \Eprint
  {http://arxiv.org/abs/2204.10171} {arXiv:2204.10171 [nucl-ex]} \BibitemShut
  {NoStop}%
\bibitem [{\citenamefont {Zhang}\ \emph {et~al.}(2024)\citenamefont {Zhang},
  \citenamefont {Huang}, \citenamefont {Becattini},\ and\ \citenamefont
  {Sheng}}]{Zhang:2024mhs}%
  \BibitemOpen
  \bibfield  {author} {\bibinfo {author} {\bibfnamefont {Z.-H.}\ \bibnamefont
  {Zhang}}, \bibinfo {author} {\bibfnamefont {X.-G.}\ \bibnamefont {Huang}},
  \bibinfo {author} {\bibfnamefont {F.}~\bibnamefont {Becattini}}, \ and\
  \bibinfo {author} {\bibfnamefont {X.-L.}\ \bibnamefont {Sheng}},\ }\href@noop
  {} {\  (\bibinfo {year} {2024})},\ \Eprint {http://arxiv.org/abs/2412.19416}
  {arXiv:2412.19416 [hep-ph]} \BibitemShut {NoStop}%
\bibitem [{\citenamefont {Kumar}\ \emph {et~al.}(2024)\citenamefont {Kumar},
  \citenamefont {Yang},\ and\ \citenamefont {Gubler}}]{Kumar:2023ojl}%
  \BibitemOpen
  \bibfield  {author} {\bibinfo {author} {\bibfnamefont {A.}~\bibnamefont
  {Kumar}}, \bibinfo {author} {\bibfnamefont {D.-L.}\ \bibnamefont {Yang}}, \
  and\ \bibinfo {author} {\bibfnamefont {P.}~\bibnamefont {Gubler}},\ }\href
  {\doibase 10.1103/PhysRevD.109.054038} {\bibfield  {journal} {\bibinfo
  {journal} {Phys. Rev. D}\ }\textbf {\bibinfo {volume} {109}},\ \bibinfo
  {pages} {054038} (\bibinfo {year} {2024})},\ \Eprint
  {http://arxiv.org/abs/2312.16900} {arXiv:2312.16900 [nucl-th]} \BibitemShut
  {NoStop}%
\bibitem [{\citenamefont {Wagner}\ \emph
  {et~al.}(2023{\natexlab{a}})\citenamefont {Wagner}, \citenamefont
  {Weickgenannt},\ and\ \citenamefont {Speranza}}]{Wagner:2022gza}%
  \BibitemOpen
  \bibfield  {author} {\bibinfo {author} {\bibfnamefont {D.}~\bibnamefont
  {Wagner}}, \bibinfo {author} {\bibfnamefont {N.}~\bibnamefont
  {Weickgenannt}}, \ and\ \bibinfo {author} {\bibfnamefont {E.}~\bibnamefont
  {Speranza}},\ }\href {\doibase 10.1103/PhysRevResearch.5.013187} {\bibfield
  {journal} {\bibinfo  {journal} {Phys. Rev. Res.}\ }\textbf {\bibinfo {volume}
  {5}},\ \bibinfo {pages} {013187} (\bibinfo {year} {2023}{\natexlab{a}})},\
  \Eprint {http://arxiv.org/abs/2207.01111} {arXiv:2207.01111 [nucl-th]}
  \BibitemShut {NoStop}%
\bibitem [{\citenamefont {Wagner}\ \emph
  {et~al.}(2023{\natexlab{b}})\citenamefont {Wagner}, \citenamefont
  {Weickgenannt},\ and\ \citenamefont {Speranza}}]{Wagner:2023cct}%
  \BibitemOpen
  \bibfield  {author} {\bibinfo {author} {\bibfnamefont {D.}~\bibnamefont
  {Wagner}}, \bibinfo {author} {\bibfnamefont {N.}~\bibnamefont
  {Weickgenannt}}, \ and\ \bibinfo {author} {\bibfnamefont {E.}~\bibnamefont
  {Speranza}},\ }\href {\doibase 10.1103/PhysRevD.108.116017} {\bibfield
  {journal} {\bibinfo  {journal} {Phys. Rev. D}\ }\textbf {\bibinfo {volume}
  {108}},\ \bibinfo {pages} {116017} (\bibinfo {year} {2023}{\natexlab{b}})},\
  \Eprint {http://arxiv.org/abs/2306.05936} {arXiv:2306.05936 [nucl-th]}
  \BibitemShut {NoStop}%
\bibitem [{\citenamefont {Gon{\c{c}}alves}\ \emph {et~al.}(2025)\citenamefont
  {Gon{\c{c}}alves}, \citenamefont {Torrieri},\ and\ \citenamefont
  {Ryblewski}}]{Goncalves:2024xzo}%
  \BibitemOpen
  \bibfield  {author} {\bibinfo {author} {\bibfnamefont {K.~J.}\ \bibnamefont
  {Gon{\c{c}}alves}}, \bibinfo {author} {\bibfnamefont {G.}~\bibnamefont
  {Torrieri}}, \ and\ \bibinfo {author} {\bibfnamefont {R.}~\bibnamefont
  {Ryblewski}},\ }\href {\doibase 10.1103/gsv6-zydk} {\bibfield  {journal}
  {\bibinfo  {journal} {Phys. Rev. C}\ }\textbf {\bibinfo {volume} {112}},\
  \bibinfo {pages} {014901} (\bibinfo {year} {2025})},\ \Eprint
  {http://arxiv.org/abs/2410.16448} {arXiv:2410.16448 [hep-ph]} \BibitemShut
  {NoStop}%
\bibitem [{\citenamefont {De~Moura}\ \emph {et~al.}(2023)\citenamefont
  {De~Moura}, \citenamefont {Goncalves},\ and\ \citenamefont
  {Torrieri}}]{DeMoura:2023jzz}%
  \BibitemOpen
  \bibfield  {author} {\bibinfo {author} {\bibfnamefont {P.~H.}\ \bibnamefont
  {De~Moura}}, \bibinfo {author} {\bibfnamefont {K.~J.}\ \bibnamefont
  {Goncalves}}, \ and\ \bibinfo {author} {\bibfnamefont {G.}~\bibnamefont
  {Torrieri}},\ }\href {\doibase 10.1103/PhysRevD.108.034032} {\bibfield
  {journal} {\bibinfo  {journal} {Phys. Rev. D}\ }\textbf {\bibinfo {volume}
  {108}},\ \bibinfo {pages} {034032} (\bibinfo {year} {2023})},\ \Eprint
  {http://arxiv.org/abs/2305.02985} {arXiv:2305.02985 [hep-ph]} \BibitemShut
  {NoStop}%
\bibitem [{\citenamefont {Yang}\ \emph {et~al.}(2018)\citenamefont {Yang},
  \citenamefont {Fang}, \citenamefont {Wang},\ and\ \citenamefont
  {Wang}}]{Yang:2017sdk}%
  \BibitemOpen
  \bibfield  {author} {\bibinfo {author} {\bibfnamefont {Y.-G.}\ \bibnamefont
  {Yang}}, \bibinfo {author} {\bibfnamefont {R.-H.}\ \bibnamefont {Fang}},
  \bibinfo {author} {\bibfnamefont {Q.}~\bibnamefont {Wang}}, \ and\ \bibinfo
  {author} {\bibfnamefont {X.-N.}\ \bibnamefont {Wang}},\ }\href {\doibase
  10.1103/PhysRevC.97.034917} {\bibfield  {journal} {\bibinfo  {journal} {Phys.
  Rev. C}\ }\textbf {\bibinfo {volume} {97}},\ \bibinfo {pages} {034917}
  (\bibinfo {year} {2018})},\ \Eprint {http://arxiv.org/abs/1711.06008}
  {arXiv:1711.06008 [nucl-th]} \BibitemShut {NoStop}%
\bibitem [{\citenamefont {Sheng}\ \emph {et~al.}(2020)\citenamefont {Sheng},
  \citenamefont {Oliva},\ and\ \citenamefont {Wang}}]{Sheng:2019kmk}%
  \BibitemOpen
  \bibfield  {author} {\bibinfo {author} {\bibfnamefont {X.-L.}\ \bibnamefont
  {Sheng}}, \bibinfo {author} {\bibfnamefont {L.}~\bibnamefont {Oliva}}, \ and\
  \bibinfo {author} {\bibfnamefont {Q.}~\bibnamefont {Wang}},\ }\href {\doibase
  10.1103/PhysRevD.101.096005} {\bibfield  {journal} {\bibinfo  {journal}
  {Phys. Rev. D}\ }\textbf {\bibinfo {volume} {101}},\ \bibinfo {pages}
  {096005} (\bibinfo {year} {2020})},\ \bibinfo {note} {[Erratum: Phys.Rev.D
  105, 099903 (2022)]},\ \Eprint {http://arxiv.org/abs/1910.13684}
  {arXiv:1910.13684 [nucl-th]} \BibitemShut {NoStop}%
\bibitem [{\citenamefont {Xia}\ \emph {et~al.}(2021)\citenamefont {Xia},
  \citenamefont {Li}, \citenamefont {Huang},\ and\ \citenamefont
  {Zhong~Huang}}]{Xia:2020tyd}%
  \BibitemOpen
  \bibfield  {author} {\bibinfo {author} {\bibfnamefont {X.-L.}\ \bibnamefont
  {Xia}}, \bibinfo {author} {\bibfnamefont {H.}~\bibnamefont {Li}}, \bibinfo
  {author} {\bibfnamefont {X.-G.}\ \bibnamefont {Huang}}, \ and\ \bibinfo
  {author} {\bibfnamefont {H.}~\bibnamefont {Zhong~Huang}},\ }\href {\doibase
  10.1016/j.physletb.2021.136325} {\bibfield  {journal} {\bibinfo  {journal}
  {Phys. Lett. B}\ }\textbf {\bibinfo {volume} {817}},\ \bibinfo {pages}
  {136325} (\bibinfo {year} {2021})},\ \Eprint
  {http://arxiv.org/abs/2010.01474} {arXiv:2010.01474 [nucl-th]} \BibitemShut
  {NoStop}%
\bibitem [{\citenamefont {Sheng}\ \emph
  {et~al.}(2024{\natexlab{a}})\citenamefont {Sheng}, \citenamefont {Yang},
  \citenamefont {Zou},\ and\ \citenamefont {Hou}}]{Sheng:2022ssp}%
  \BibitemOpen
  \bibfield  {author} {\bibinfo {author} {\bibfnamefont {X.-L.}\ \bibnamefont
  {Sheng}}, \bibinfo {author} {\bibfnamefont {S.-Y.}\ \bibnamefont {Yang}},
  \bibinfo {author} {\bibfnamefont {Y.-L.}\ \bibnamefont {Zou}}, \ and\
  \bibinfo {author} {\bibfnamefont {D.}~\bibnamefont {Hou}},\ }\href {\doibase
  10.1140/epjc/s10052-024-12643-7} {\bibfield  {journal} {\bibinfo  {journal}
  {Eur. Phys. J. C}\ }\textbf {\bibinfo {volume} {84}},\ \bibinfo {pages} {299}
  (\bibinfo {year} {2024}{\natexlab{a}})},\ \Eprint
  {http://arxiv.org/abs/2209.01872} {arXiv:2209.01872 [nucl-th]} \BibitemShut
  {NoStop}%
\bibitem [{\citenamefont {Li}\ \emph {et~al.}(2023)\citenamefont {Li},
  \citenamefont {Xia}, \citenamefont {Huang},\ and\ \citenamefont
  {Huang}}]{Li:2023tsf}%
  \BibitemOpen
  \bibfield  {author} {\bibinfo {author} {\bibfnamefont {H.}~\bibnamefont
  {Li}}, \bibinfo {author} {\bibfnamefont {X.-L.}\ \bibnamefont {Xia}},
  \bibinfo {author} {\bibfnamefont {X.-G.}\ \bibnamefont {Huang}}, \ and\
  \bibinfo {author} {\bibfnamefont {H.~Z.}\ \bibnamefont {Huang}},\ }\href
  {\doibase 10.1103/PhysRevC.108.044902} {\bibfield  {journal} {\bibinfo
  {journal} {Phys. Rev. C}\ }\textbf {\bibinfo {volume} {108}},\ \bibinfo
  {pages} {044902} (\bibinfo {year} {2023})},\ \Eprint
  {http://arxiv.org/abs/2306.02829} {arXiv:2306.02829 [nucl-th]} \BibitemShut
  {NoStop}%
\bibitem [{\citenamefont {Sheng}\ \emph
  {et~al.}(2023{\natexlab{a}})\citenamefont {Sheng}, \citenamefont {Oliva},
  \citenamefont {Liang}, \citenamefont {Wang},\ and\ \citenamefont
  {Wang}}]{Sheng:2022wsy}%
  \BibitemOpen
  \bibfield  {author} {\bibinfo {author} {\bibfnamefont {X.-L.}\ \bibnamefont
  {Sheng}}, \bibinfo {author} {\bibfnamefont {L.}~\bibnamefont {Oliva}},
  \bibinfo {author} {\bibfnamefont {Z.-T.}\ \bibnamefont {Liang}}, \bibinfo
  {author} {\bibfnamefont {Q.}~\bibnamefont {Wang}}, \ and\ \bibinfo {author}
  {\bibfnamefont {X.-N.}\ \bibnamefont {Wang}},\ }\href {\doibase
  10.1103/PhysRevLett.131.042304} {\bibfield  {journal} {\bibinfo  {journal}
  {Phys. Rev. Lett.}\ }\textbf {\bibinfo {volume} {131}},\ \bibinfo {pages}
  {042304} (\bibinfo {year} {2023}{\natexlab{a}})},\ \Eprint
  {http://arxiv.org/abs/2205.15689} {arXiv:2205.15689 [nucl-th]} \BibitemShut
  {NoStop}%
\bibitem [{\citenamefont {Sheng}\ \emph
  {et~al.}(2023{\natexlab{b}})\citenamefont {Sheng}, \citenamefont {Pu},\ and\
  \citenamefont {Wang}}]{Sheng:2023urn}%
  \BibitemOpen
  \bibfield  {author} {\bibinfo {author} {\bibfnamefont {X.-L.}\ \bibnamefont
  {Sheng}}, \bibinfo {author} {\bibfnamefont {S.}~\bibnamefont {Pu}}, \ and\
  \bibinfo {author} {\bibfnamefont {Q.}~\bibnamefont {Wang}},\ }\href {\doibase
  10.1103/PhysRevC.108.054902} {\bibfield  {journal} {\bibinfo  {journal}
  {Phys. Rev. C}\ }\textbf {\bibinfo {volume} {108}},\ \bibinfo {pages}
  {054902} (\bibinfo {year} {2023}{\natexlab{b}})},\ \Eprint
  {http://arxiv.org/abs/2308.14038} {arXiv:2308.14038 [nucl-th]} \BibitemShut
  {NoStop}%
\bibitem [{\citenamefont {Kumar}\ \emph
  {et~al.}(2023{\natexlab{a}})\citenamefont {Kumar}, \citenamefont
  {M{\"u}ller},\ and\ \citenamefont {Yang}}]{Kumar:2022ylt}%
  \BibitemOpen
  \bibfield  {author} {\bibinfo {author} {\bibfnamefont {A.}~\bibnamefont
  {Kumar}}, \bibinfo {author} {\bibfnamefont {B.}~\bibnamefont {M{\"u}ller}}, \
  and\ \bibinfo {author} {\bibfnamefont {D.-L.}\ \bibnamefont {Yang}},\ }\href
  {\doibase 10.1103/PhysRevD.107.076025} {\bibfield  {journal} {\bibinfo
  {journal} {Phys. Rev. D}\ }\textbf {\bibinfo {volume} {107}},\ \bibinfo
  {pages} {076025} (\bibinfo {year} {2023}{\natexlab{a}})},\ \Eprint
  {http://arxiv.org/abs/2212.13354} {arXiv:2212.13354 [nucl-th]} \BibitemShut
  {NoStop}%
\bibitem [{\citenamefont {M\"uller}\ and\ \citenamefont
  {Yang}(2022)}]{Muller:2021hpe}%
  \BibitemOpen
  \bibfield  {author} {\bibinfo {author} {\bibfnamefont {B.}~\bibnamefont
  {M\"uller}}\ and\ \bibinfo {author} {\bibfnamefont {D.-L.}\ \bibnamefont
  {Yang}},\ }\href {\doibase 10.1103/PhysRevD.105.L011901} {\bibfield
  {journal} {\bibinfo  {journal} {Phys. Rev. D}\ }\textbf {\bibinfo {volume}
  {105}},\ \bibinfo {pages} {L011901} (\bibinfo {year} {2022})},\ \bibinfo
  {note} {[Erratum: Phys.Rev.D 106, 039904 (2022)]},\ \Eprint
  {http://arxiv.org/abs/2110.15630} {arXiv:2110.15630 [nucl-th]} \BibitemShut
  {NoStop}%
\bibitem [{\citenamefont {Kumar}\ \emph
  {et~al.}(2023{\natexlab{b}})\citenamefont {Kumar}, \citenamefont
  {M{\"u}ller},\ and\ \citenamefont {Yang}}]{Kumar:2023ghs}%
  \BibitemOpen
  \bibfield  {author} {\bibinfo {author} {\bibfnamefont {A.}~\bibnamefont
  {Kumar}}, \bibinfo {author} {\bibfnamefont {B.}~\bibnamefont {M{\"u}ller}}, \
  and\ \bibinfo {author} {\bibfnamefont {D.-L.}\ \bibnamefont {Yang}},\ }\href
  {\doibase 10.1103/PhysRevD.108.016020} {\bibfield  {journal} {\bibinfo
  {journal} {Phys. Rev. D}\ }\textbf {\bibinfo {volume} {108}},\ \bibinfo
  {pages} {016020} (\bibinfo {year} {2023}{\natexlab{b}})},\ \Eprint
  {http://arxiv.org/abs/2304.04181} {arXiv:2304.04181 [nucl-th]} \BibitemShut
  {NoStop}%
\bibitem [{\citenamefont {Yang}(2025)}]{Yang:2024qpy}%
  \BibitemOpen
  \bibfield  {author} {\bibinfo {author} {\bibfnamefont {D.-L.}\ \bibnamefont
  {Yang}},\ }\href {\doibase 10.1103/PhysRevD.111.056005} {\bibfield  {journal}
  {\bibinfo  {journal} {Phys. Rev. D}\ }\textbf {\bibinfo {volume} {111}},\
  \bibinfo {pages} {056005} (\bibinfo {year} {2025})},\ \Eprint
  {http://arxiv.org/abs/2411.14822} {arXiv:2411.14822 [nucl-th]} \BibitemShut
  {NoStop}%
\bibitem [{\citenamefont {Liang}\ and\ \citenamefont
  {Lin}(2025)}]{Liang:2025hxw}%
  \BibitemOpen
  \bibfield  {author} {\bibinfo {author} {\bibfnamefont {Y.}~\bibnamefont
  {Liang}}\ and\ \bibinfo {author} {\bibfnamefont {S.}~\bibnamefont {Lin}},\
  }\href {\doibase 10.1088/1674-1137/adcc8c} {\bibfield  {journal} {\bibinfo
  {journal} {Chin. Phys. C}\ }\textbf {\bibinfo {volume} {49}},\ \bibinfo
  {pages} {084105} (\bibinfo {year} {2025})},\ \Eprint
  {http://arxiv.org/abs/2502.05866} {arXiv:2502.05866 [hep-ph]} \BibitemShut
  {NoStop}%
\bibitem [{\citenamefont {Chen}\ and\ \citenamefont
  {Lin}(2025)}]{Chen:2025mrf}%
  \BibitemOpen
  \bibfield  {author} {\bibinfo {author} {\bibfnamefont {Z.}~\bibnamefont
  {Chen}}\ and\ \bibinfo {author} {\bibfnamefont {S.}~\bibnamefont {Lin}},\
  }\href {\doibase 10.1103/PhysRevD.111.074002} {\bibfield  {journal} {\bibinfo
   {journal} {Phys. Rev. D}\ }\textbf {\bibinfo {volume} {111}},\ \bibinfo
  {pages} {074002} (\bibinfo {year} {2025})},\ \Eprint
  {http://arxiv.org/abs/2501.16596} {arXiv:2501.16596 [hep-ph]} \BibitemShut
  {NoStop}%
\bibitem [{\citenamefont {Li}\ and\ \citenamefont {Liu}(2022)}]{Li:2022vmb}%
  \BibitemOpen
  \bibfield  {author} {\bibinfo {author} {\bibfnamefont {F.}~\bibnamefont
  {Li}}\ and\ \bibinfo {author} {\bibfnamefont {S.~Y.~F.}\ \bibnamefont
  {Liu}},\ }\href@noop {} {\  (\bibinfo {year} {2022})},\ \Eprint
  {http://arxiv.org/abs/2206.11890} {arXiv:2206.11890 [nucl-th]} \BibitemShut
  {NoStop}%
\bibitem [{\citenamefont {Grossi}\ \emph {et~al.}(2025)\citenamefont {Grossi},
  \citenamefont {Palermo},\ and\ \citenamefont {Zahed}}]{Grossi:2024pyh}%
  \BibitemOpen
  \bibfield  {author} {\bibinfo {author} {\bibfnamefont {E.}~\bibnamefont
  {Grossi}}, \bibinfo {author} {\bibfnamefont {A.}~\bibnamefont {Palermo}}, \
  and\ \bibinfo {author} {\bibfnamefont {I.}~\bibnamefont {Zahed}},\ }\href
  {\doibase 10.1103/PhysRevC.111.014914} {\bibfield  {journal} {\bibinfo
  {journal} {Phys. Rev. C}\ }\textbf {\bibinfo {volume} {111}},\ \bibinfo
  {pages} {014914} (\bibinfo {year} {2025})},\ \Eprint
  {http://arxiv.org/abs/2407.10524} {arXiv:2407.10524 [nucl-th]} \BibitemShut
  {NoStop}%
\bibitem [{\citenamefont {Li}(2024)}]{Li:2024qae}%
  \BibitemOpen
  \bibfield  {author} {\bibinfo {author} {\bibfnamefont {F.}~\bibnamefont
  {Li}},\ }\href {\doibase 10.1103/PhysRevC.109.064916} {\bibfield  {journal}
  {\bibinfo  {journal} {Phys. Rev. C}\ }\textbf {\bibinfo {volume} {109}},\
  \bibinfo {pages} {064916} (\bibinfo {year} {2024})},\ \Eprint
  {http://arxiv.org/abs/2404.02860} {arXiv:2404.02860 [nucl-th]} \BibitemShut
  {NoStop}%
\bibitem [{\citenamefont {Park}\ \emph {et~al.}(2023)\citenamefont {Park},
  \citenamefont {Sako}, \citenamefont {Aoki}, \citenamefont {Gubler},\ and\
  \citenamefont {Lee}}]{Park:2022ayr}%
  \BibitemOpen
  \bibfield  {author} {\bibinfo {author} {\bibfnamefont {I.~W.}\ \bibnamefont
  {Park}}, \bibinfo {author} {\bibfnamefont {H.}~\bibnamefont {Sako}}, \bibinfo
  {author} {\bibfnamefont {K.}~\bibnamefont {Aoki}}, \bibinfo {author}
  {\bibfnamefont {P.}~\bibnamefont {Gubler}}, \ and\ \bibinfo {author}
  {\bibfnamefont {S.~H.}\ \bibnamefont {Lee}},\ }\href {\doibase
  10.1103/PhysRevD.107.074033} {\bibfield  {journal} {\bibinfo  {journal}
  {Phys. Rev. D}\ }\textbf {\bibinfo {volume} {107}},\ \bibinfo {pages}
  {074033} (\bibinfo {year} {2023})},\ \Eprint
  {http://arxiv.org/abs/2211.16949} {arXiv:2211.16949 [hep-ph]} \BibitemShut
  {NoStop}%
\bibitem [{\citenamefont {Sun}\ \emph {et~al.}(2025)\citenamefont {Sun},
  \citenamefont {Li},\ and\ \citenamefont {Liu}}]{Sun:2025ror}%
  \BibitemOpen
  \bibfield  {author} {\bibinfo {author} {\bibfnamefont {Z.-Y.}\ \bibnamefont
  {Sun}}, \bibinfo {author} {\bibfnamefont {Y.-Y.}\ \bibnamefont {Li}}, \ and\
  \bibinfo {author} {\bibfnamefont {S.~Y.~F.}\ \bibnamefont {Liu}},\
  }\href@noop {} {\  (\bibinfo {year} {2025})},\ \Eprint
  {http://arxiv.org/abs/2503.13408} {arXiv:2503.13408 [nucl-th]} \BibitemShut
  {NoStop}%
\bibitem [{\citenamefont {Yin}\ \emph {et~al.}(2024)\citenamefont {Yin},
  \citenamefont {Dong}, \citenamefont {Pang}, \citenamefont {Pu},\ and\
  \citenamefont {Wang}}]{Yin:2024dnu}%
  \BibitemOpen
  \bibfield  {author} {\bibinfo {author} {\bibfnamefont {Y.-L.}\ \bibnamefont
  {Yin}}, \bibinfo {author} {\bibfnamefont {W.-B.}\ \bibnamefont {Dong}},
  \bibinfo {author} {\bibfnamefont {J.-Y.}\ \bibnamefont {Pang}}, \bibinfo
  {author} {\bibfnamefont {S.}~\bibnamefont {Pu}}, \ and\ \bibinfo {author}
  {\bibfnamefont {Q.}~\bibnamefont {Wang}},\ }\href@noop {} {\  (\bibinfo
  {year} {2024})},\ \Eprint {http://arxiv.org/abs/2402.03672} {arXiv:2402.03672
  [nucl-th]} \BibitemShut {NoStop}%
\bibitem [{\citenamefont {Zhao}\ \emph {et~al.}(2024)\citenamefont {Zhao},
  \citenamefont {Sheng}, \citenamefont {Li},\ and\ \citenamefont
  {Hou}}]{Zhao:2024ipr}%
  \BibitemOpen
  \bibfield  {author} {\bibinfo {author} {\bibfnamefont {Y.-Q.}\ \bibnamefont
  {Zhao}}, \bibinfo {author} {\bibfnamefont {X.-L.}\ \bibnamefont {Sheng}},
  \bibinfo {author} {\bibfnamefont {S.-W.}\ \bibnamefont {Li}}, \ and\ \bibinfo
  {author} {\bibfnamefont {D.}~\bibnamefont {Hou}},\ }\href {\doibase
  10.1007/JHEP08(2024)070} {\bibfield  {journal} {\bibinfo  {journal} {JHEP}\
  }\textbf {\bibinfo {volume} {08}},\ \bibinfo {pages} {070} (\bibinfo {year}
  {2024})},\ \Eprint {http://arxiv.org/abs/2403.07468} {arXiv:2403.07468
  [hep-ph]} \BibitemShut {NoStop}%
\bibitem [{\citenamefont {Sheng}\ \emph
  {et~al.}(2024{\natexlab{b}})\citenamefont {Sheng}, \citenamefont {Zhao},
  \citenamefont {Li}, \citenamefont {Becattini},\ and\ \citenamefont
  {Hou}}]{Sheng:2024kgg}%
  \BibitemOpen
  \bibfield  {author} {\bibinfo {author} {\bibfnamefont {X.-L.}\ \bibnamefont
  {Sheng}}, \bibinfo {author} {\bibfnamefont {Y.-Q.}\ \bibnamefont {Zhao}},
  \bibinfo {author} {\bibfnamefont {S.-W.}\ \bibnamefont {Li}}, \bibinfo
  {author} {\bibfnamefont {F.}~\bibnamefont {Becattini}}, \ and\ \bibinfo
  {author} {\bibfnamefont {D.}~\bibnamefont {Hou}},\ }\href@noop {} {\
  (\bibinfo {year} {2024}{\natexlab{b}})},\ \Eprint
  {http://arxiv.org/abs/2403.07522} {arXiv:2403.07522 [hep-ph]} \BibitemShut
  {NoStop}%
\bibitem [{\citenamefont {Yang}\ and\ \citenamefont
  {Yao}(2024)}]{Yang:2024ejk}%
  \BibitemOpen
  \bibfield  {author} {\bibinfo {author} {\bibfnamefont {D.-L.}\ \bibnamefont
  {Yang}}\ and\ \bibinfo {author} {\bibfnamefont {X.}~\bibnamefont {Yao}},\
  }\href@noop {} {\  (\bibinfo {year} {2024})},\ \Eprint
  {http://arxiv.org/abs/2405.20280} {arXiv:2405.20280 [hep-ph]} \BibitemShut
  {NoStop}%
\bibitem [{\citenamefont {Yamagishi}\ and\ \citenamefont
  {Zahed}(1996)}]{Yamagishi:1995kr}%
  \BibitemOpen
  \bibfield  {author} {\bibinfo {author} {\bibfnamefont {H.}~\bibnamefont
  {Yamagishi}}\ and\ \bibinfo {author} {\bibfnamefont {I.}~\bibnamefont
  {Zahed}},\ }\href {\doibase 10.1006/aphy.1996.0045} {\bibfield  {journal}
  {\bibinfo  {journal} {Annals Phys.}\ }\textbf {\bibinfo {volume} {247}},\
  \bibinfo {pages} {292} (\bibinfo {year} {1996})},\ \Eprint
  {http://arxiv.org/abs/hep-ph/9503413} {arXiv:hep-ph/9503413} \BibitemShut
  {NoStop}%
\bibitem [{\citenamefont {Kamano}(2010)}]{Kamano:2009st}%
  \BibitemOpen
  \bibfield  {author} {\bibinfo {author} {\bibfnamefont {H.}~\bibnamefont
  {Kamano}},\ }\href {\doibase 10.1103/PhysRevD.81.076004} {\bibfield
  {journal} {\bibinfo  {journal} {Phys. Rev. D}\ }\textbf {\bibinfo {volume}
  {81}},\ \bibinfo {pages} {076004} (\bibinfo {year} {2010})},\ \Eprint
  {http://arxiv.org/abs/0909.1606} {arXiv:0909.1606 [hep-ph]} \BibitemShut
  {NoStop}%
\bibitem [{\citenamefont {Steele}\ \emph {et~al.}(1996)\citenamefont {Steele},
  \citenamefont {Yamagishi},\ and\ \citenamefont {Zahed}}]{Steele:1996su}%
  \BibitemOpen
  \bibfield  {author} {\bibinfo {author} {\bibfnamefont {J.~V.}\ \bibnamefont
  {Steele}}, \bibinfo {author} {\bibfnamefont {H.}~\bibnamefont {Yamagishi}}, \
  and\ \bibinfo {author} {\bibfnamefont {I.}~\bibnamefont {Zahed}},\ }\href
  {\doibase 10.1016/0370-2693(96)00802-7} {\bibfield  {journal} {\bibinfo
  {journal} {Phys. Lett. B}\ }\textbf {\bibinfo {volume} {384}},\ \bibinfo
  {pages} {255} (\bibinfo {year} {1996})},\ \Eprint
  {http://arxiv.org/abs/hep-ph/9603290} {arXiv:hep-ph/9603290} \BibitemShut
  {NoStop}%
\bibitem [{\citenamefont {Braghin}(2022)}]{Braghin:2021hmr}%
  \BibitemOpen
  \bibfield  {author} {\bibinfo {author} {\bibfnamefont {F.~L.}\ \bibnamefont
  {Braghin}},\ }\href {\doibase 10.1088/1361-6471/ac4d79} {\bibfield  {journal}
  {\bibinfo  {journal} {J. Phys. G}\ }\textbf {\bibinfo {volume} {49}},\
  \bibinfo {pages} {055101} (\bibinfo {year} {2022})},\ \bibinfo {note}
  {[Erratum: J.Phys.G 50, 119501 (2023)]},\ \Eprint
  {http://arxiv.org/abs/2108.02748} {arXiv:2108.02748 [hep-ph]} \BibitemShut
  {NoStop}%
\bibitem [{\citenamefont {Liu}\ and\ \citenamefont
  {Zahed}(2017)}]{Liu:2017fib}%
  \BibitemOpen
  \bibfield  {author} {\bibinfo {author} {\bibfnamefont {Y.}~\bibnamefont
  {Liu}}\ and\ \bibinfo {author} {\bibfnamefont {I.}~\bibnamefont {Zahed}},\
  }\href {\doibase 10.1103/PhysRevD.96.116021} {\bibfield  {journal} {\bibinfo
  {journal} {Phys. Rev. D}\ }\textbf {\bibinfo {volume} {96}},\ \bibinfo
  {pages} {116021} (\bibinfo {year} {2017})},\ \Eprint
  {http://arxiv.org/abs/1707.08523} {arXiv:1707.08523 [hep-ph]} \BibitemShut
  {NoStop}%
\bibitem [{\citenamefont {Group}(2022)}]{pdg}%
  \BibitemOpen
  \bibfield  {author} {\bibinfo {author} {\bibfnamefont {P.~D.}\ \bibnamefont
  {Group}},\ }\href {\doibase 10.1093/ptep/ptac097} {\bibfield  {journal}
  {\bibinfo  {journal} {Progress of Theoretical and Experimental Physics}\
  }\textbf {\bibinfo {volume} {2022}},\ \bibinfo {pages} {083C01} (\bibinfo
  {year} {2022})},\ \Eprint
  {http://arxiv.org/abs/https://academic.oup.com/ptep/article-pdf/2022/8/083C01/49175539/ptac097.pdf}
  {https://academic.oup.com/ptep/article-pdf/2022/8/083C01/49175539/ptac097.pdf}
  \BibitemShut {NoStop}%
\bibitem [{\citenamefont {Stoks}\ and\ \citenamefont
  {Rijken}(1999)}]{PhysRevC.59.3009}%
  \BibitemOpen
  \bibfield  {author} {\bibinfo {author} {\bibfnamefont {V.~G.~J.}\
  \bibnamefont {Stoks}}\ and\ \bibinfo {author} {\bibfnamefont {T.~A.}\
  \bibnamefont {Rijken}},\ }\href {\doibase 10.1103/PhysRevC.59.3009}
  {\bibfield  {journal} {\bibinfo  {journal} {Phys. Rev. C}\ }\textbf {\bibinfo
  {volume} {59}},\ \bibinfo {pages} {3009} (\bibinfo {year}
  {1999})}\BibitemShut {NoStop}%
\bibitem [{\citenamefont {Rijken}\ \emph {et~al.}(1999)\citenamefont {Rijken},
  \citenamefont {Stoks},\ and\ \citenamefont {Yamamoto}}]{PhysRevC.59.21}%
  \BibitemOpen
  \bibfield  {author} {\bibinfo {author} {\bibfnamefont {T.~A.}\ \bibnamefont
  {Rijken}}, \bibinfo {author} {\bibfnamefont {V.~G.~J.}\ \bibnamefont
  {Stoks}}, \ and\ \bibinfo {author} {\bibfnamefont {Y.}~\bibnamefont
  {Yamamoto}},\ }\href {\doibase 10.1103/PhysRevC.59.21} {\bibfield  {journal}
  {\bibinfo  {journal} {Phys. Rev. C}\ }\textbf {\bibinfo {volume} {59}},\
  \bibinfo {pages} {21} (\bibinfo {year} {1999})}\BibitemShut {NoStop}%
\bibitem [{\citenamefont {Floerchinger}\ \emph {et~al.}(2019)\citenamefont
  {Floerchinger}, \citenamefont {Grossi},\ and\ \citenamefont
  {Lion}}]{Floerchinger:2018pje}%
  \BibitemOpen
  \bibfield  {author} {\bibinfo {author} {\bibfnamefont {S.}~\bibnamefont
  {Floerchinger}}, \bibinfo {author} {\bibfnamefont {E.}~\bibnamefont
  {Grossi}}, \ and\ \bibinfo {author} {\bibfnamefont {J.}~\bibnamefont
  {Lion}},\ }\href {\doibase 10.1103/PhysRevC.100.014905} {\bibfield  {journal}
  {\bibinfo  {journal} {Phys. Rev. C}\ }\textbf {\bibinfo {volume} {100}},\
  \bibinfo {pages} {014905} (\bibinfo {year} {2019})},\ \Eprint
  {http://arxiv.org/abs/1811.01870} {arXiv:1811.01870 [nucl-th]} \BibitemShut
  {NoStop}%
\bibitem [{\citenamefont {Devetak}\ \emph {et~al.}(2020)\citenamefont
  {Devetak}, \citenamefont {Dubla}, \citenamefont {Floerchinger}, \citenamefont
  {Grossi}, \citenamefont {Masciocchi}, \citenamefont {Mazeliauskas},\ and\
  \citenamefont {Selyuzhenkov}}]{Devetak:2019lsk}%
  \BibitemOpen
  \bibfield  {author} {\bibinfo {author} {\bibfnamefont {D.}~\bibnamefont
  {Devetak}}, \bibinfo {author} {\bibfnamefont {A.}~\bibnamefont {Dubla}},
  \bibinfo {author} {\bibfnamefont {S.}~\bibnamefont {Floerchinger}}, \bibinfo
  {author} {\bibfnamefont {E.}~\bibnamefont {Grossi}}, \bibinfo {author}
  {\bibfnamefont {S.}~\bibnamefont {Masciocchi}}, \bibinfo {author}
  {\bibfnamefont {A.}~\bibnamefont {Mazeliauskas}}, \ and\ \bibinfo {author}
  {\bibfnamefont {I.}~\bibnamefont {Selyuzhenkov}},\ }\href {\doibase
  10.1007/JHEP06(2020)044} {\bibfield  {journal} {\bibinfo  {journal} {JHEP}\
  }\textbf {\bibinfo {volume} {06}},\ \bibinfo {pages} {044} (\bibinfo {year}
  {2020})},\ \Eprint {http://arxiv.org/abs/1909.10485} {arXiv:1909.10485
  [hep-ph]} \BibitemShut {NoStop}%
\bibitem [{\citenamefont {Capellino}\ \emph {et~al.}(2023)\citenamefont
  {Capellino}, \citenamefont {Dubla}, \citenamefont {Floerchinger},
  \citenamefont {Grossi}, \citenamefont {Kirchner},\ and\ \citenamefont
  {Masciocchi}}]{Capellino:2023cxe}%
  \BibitemOpen
  \bibfield  {author} {\bibinfo {author} {\bibfnamefont {F.}~\bibnamefont
  {Capellino}}, \bibinfo {author} {\bibfnamefont {A.}~\bibnamefont {Dubla}},
  \bibinfo {author} {\bibfnamefont {S.}~\bibnamefont {Floerchinger}}, \bibinfo
  {author} {\bibfnamefont {E.}~\bibnamefont {Grossi}}, \bibinfo {author}
  {\bibfnamefont {A.}~\bibnamefont {Kirchner}}, \ and\ \bibinfo {author}
  {\bibfnamefont {S.}~\bibnamefont {Masciocchi}},\ }\href {\doibase
  10.1103/PhysRevD.108.116011} {\bibfield  {journal} {\bibinfo  {journal}
  {Phys. Rev. D}\ }\textbf {\bibinfo {volume} {108}},\ \bibinfo {pages}
  {116011} (\bibinfo {year} {2023})},\ \Eprint
  {http://arxiv.org/abs/2307.14449} {arXiv:2307.14449 [hep-ph]} \BibitemShut
  {NoStop}%
\bibitem [{\citenamefont {Adam}\ \emph {et~al.}(2020)\citenamefont {Adam} \emph
  {et~al.}}]{STAR:2019bjj}%
  \BibitemOpen
  \bibfield  {author} {\bibinfo {author} {\bibfnamefont {J.}~\bibnamefont
  {Adam}} \emph {et~al.} (\bibinfo {collaboration} {STAR}),\ }\href {\doibase
  10.1103/PhysRevC.102.034909} {\bibfield  {journal} {\bibinfo  {journal}
  {Phys. Rev. C}\ }\textbf {\bibinfo {volume} {102}},\ \bibinfo {pages}
  {034909} (\bibinfo {year} {2020})},\ \Eprint
  {http://arxiv.org/abs/1906.03732} {arXiv:1906.03732 [nucl-ex]} \BibitemShut
  {NoStop}%
\bibitem [{\citenamefont {Bollweg}\ \emph {et~al.}(2024)\citenamefont
  {Bollweg}, \citenamefont {Ding}, \citenamefont {Goswami}, \citenamefont
  {Karsch}, \citenamefont {Mukherjee}, \citenamefont {Petreczky},\ and\
  \citenamefont {Schmidt}}]{Bollweg:2024epj}%
  \BibitemOpen
  \bibfield  {author} {\bibinfo {author} {\bibfnamefont {D.}~\bibnamefont
  {Bollweg}}, \bibinfo {author} {\bibfnamefont {H.~T.}\ \bibnamefont {Ding}},
  \bibinfo {author} {\bibfnamefont {J.}~\bibnamefont {Goswami}}, \bibinfo
  {author} {\bibfnamefont {F.}~\bibnamefont {Karsch}}, \bibinfo {author}
  {\bibfnamefont {S.}~\bibnamefont {Mukherjee}}, \bibinfo {author}
  {\bibfnamefont {P.}~\bibnamefont {Petreczky}}, \ and\ \bibinfo {author}
  {\bibfnamefont {C.}~\bibnamefont {Schmidt}},\ }\href {\doibase
  10.1103/PhysRevD.110.054519} {\bibfield  {journal} {\bibinfo  {journal}
  {Phys. Rev. D}\ }\textbf {\bibinfo {volume} {110}},\ \bibinfo {pages}
  {054519} (\bibinfo {year} {2024})},\ \Eprint
  {http://arxiv.org/abs/2407.09335} {arXiv:2407.09335 [hep-lat]} \BibitemShut
  {NoStop}%
\bibitem [{\citenamefont {Moreland}\ \emph {et~al.}(2015)\citenamefont
  {Moreland}, \citenamefont {Bernhard},\ and\ \citenamefont
  {Bass}}]{Moreland:2014oya}%
  \BibitemOpen
  \bibfield  {author} {\bibinfo {author} {\bibfnamefont {J.~S.}\ \bibnamefont
  {Moreland}}, \bibinfo {author} {\bibfnamefont {J.~E.}\ \bibnamefont
  {Bernhard}}, \ and\ \bibinfo {author} {\bibfnamefont {S.~A.}\ \bibnamefont
  {Bass}},\ }\href {\doibase 10.1103/PhysRevC.92.011901} {\bibfield  {journal}
  {\bibinfo  {journal} {Phys.Rev.}\ }\textbf {\bibinfo {volume} {C92}},\
  \bibinfo {pages} {011901} (\bibinfo {year} {2015})},\ \Eprint
  {http://arxiv.org/abs/1412.4708} {arXiv:1412.4708 [nucl-th]} \BibitemShut
  {NoStop}%
\bibitem [{\citenamefont {Abelev}\ \emph {et~al.}(2009)\citenamefont {Abelev}
  \emph {et~al.}}]{STAR:2008bgi}%
  \BibitemOpen
  \bibfield  {author} {\bibinfo {author} {\bibfnamefont {B.~I.}\ \bibnamefont
  {Abelev}} \emph {et~al.} (\bibinfo {collaboration} {STAR}),\ }\href {\doibase
  10.1103/PhysRevC.79.064903} {\bibfield  {journal} {\bibinfo  {journal} {Phys.
  Rev. C}\ }\textbf {\bibinfo {volume} {79}},\ \bibinfo {pages} {064903}
  (\bibinfo {year} {2009})},\ \Eprint {http://arxiv.org/abs/0809.4737}
  {arXiv:0809.4737 [nucl-ex]} \BibitemShut {NoStop}%
\bibitem [{\citenamefont {Adamczyk}\ \emph {et~al.}(2016)\citenamefont
  {Adamczyk} \emph {et~al.}}]{STAR:2015vvs}%
  \BibitemOpen
  \bibfield  {author} {\bibinfo {author} {\bibfnamefont {L.}~\bibnamefont
  {Adamczyk}} \emph {et~al.} (\bibinfo {collaboration} {STAR}),\ }\href
  {\doibase 10.1103/PhysRevC.93.021903} {\bibfield  {journal} {\bibinfo
  {journal} {Phys. Rev. C}\ }\textbf {\bibinfo {volume} {93}},\ \bibinfo
  {pages} {021903} (\bibinfo {year} {2016})},\ \Eprint
  {http://arxiv.org/abs/1506.07605} {arXiv:1506.07605 [nucl-ex]} \BibitemShut
  {NoStop}%
\end{thebibliography}%

\end{document}